%% file: arxiv.tex
\documentclass{article}

\usepackage{arxiv}

\usepackage{graphicx}
\usepackage{dcolumn}
\usepackage{listings}

\usepackage{bm}
\usepackage[mathlines]{lineno}

\usepackage{makecell}
\usepackage{multirow}
\usepackage{hhline}
\usepackage[utf8]{inputenc}
\usepackage[T1]{fontenc}
\usepackage{mathptmx}
\usepackage{etoolbox}
\usepackage{xcolor}
\usepackage{color}
\usepackage{float}
\usepackage{hyperref}
\newcommand{\be}{\begin{equation}}
\newcommand{\ee}{\end{equation}}

\usepackage{amsmath}
\usepackage[numbers, sort&compress]{natbib}
\usepackage{hyperref}       
\usepackage{url}            
\usepackage{booktabs}       
\usepackage{amsfonts}       
\usepackage{nicefrac}       
\usepackage{microtype}      
\usepackage{lipsum}
\usepackage{graphicx}
\usepackage{hyperref} 
\hypersetup{colorlinks = true, 
	        linkcolor = blue, 
	        urlcolor = blue,
            citecolor = blue} 
\usepackage[misc]{ifsym}

\title{Modeling of turbulence kinetic energy added by wind-turbine wakes in the atmospheric boundary layer}

\author{
 Bowen Du, Jingshan Zhu, Baoliang Li, Mingwei Ge $^{\textrm{\Letter}}$, Xintao Li, and Yongqian Liu \\
 \\
  State Key Laboratory of Alternate Electrical Power System with Renewable Energy Sources,\\ North China Electric Power University, Beijing 102206, PR China \\
  \\
  \texttt{${\textrm{\Letter}}$ gemingwei@ncepu.edu.cn}\\
}

\begin{document}
\maketitle
\begin{abstract}
Accurate prediction of turbulence kinetic energy (TKE) added by wind-turbine wakes is of significant scientific value for understanding the wake recovery mechanisms. Furthermore, this physical quantity is a critical input for engineering applications such as wake velocity deficit modeling and fatigue damage assessment. In this study, we propose a novel wake-added TKE prediction model capable of accurately predict the three-dimensional spatial distribution of wake-added TKE using only basic inflow and wind turbine operation conditions as inputs. The model consists of two sub-modules: one for calculating the azimuthally-averaged wake-added TKE and the other for determining the ground effect correction function. The calculation of the azimuthally-averaged wake-added TKE is based on the analytical solution derived from the modeled azimuthally-averaged wake-added TKE budget, while the ground effect correction function is determined using a unified functional form, owing to its self-similarity. To ensure the closure of these two sub-modules, we develop methods for determining all the unknown parameters based on the large-eddy simulation (LES) calibration cases. This results in an end-to-end prediction framework, enabling direct engineering applications of the proposed model. We compared the proposed model with LES calibration data and publicly available validation datasets from the literature, which include LES and wind tunnel experiments under various inflow and turbine operating conditions. The comparison results show that the proposed model can accurately predict the spatial distribution of wake-added TKE, particularly capturing the vertical asymmetry of wake-added TKE and the streamwise evolution of the hub-height wake-added TKE profile from the dual-peak distribution to the approximately top-hat distribution. The averaged normalized mean absolute error of the proposed model across all validation datasets is only 8.13\%, demonstrating the robustness and broad applicability of the proposed model and its corresponding parameter determination methods.
\end{abstract}

\input{sections/section_1.tex}
\input{sections/section_2.tex}
\input{sections/section_3.tex}
\input{sections/section_4.tex}
\input{sections/section_5.tex}

\input{sections/section_6.tex}
\section*{Acknowledgments}
This research was supported by Beijing Natural Science Foundation (No. JQ22008) and the National Natural Science Foundation of China (No. 12172128). The authors would
like to thank Dr. Fr{\'e}d{\'e}ric Blondel for sharing the LES datasets.

\section*{Author Contributions}
Bowen Du: Conceptualization, Software, Data curation, Writing - original draft.
Jingshan Zhu: Software, Writing - review \& editing.
Baoliang Li: Methodology, Writing - review \& editing.
Mingwei Ge: Conceptualization, Methodology, Supervision, Writing - review \& editing.
Xintao Li: Writing - review \& editing.
Yongqian Liu: Supervision, Writing - review \& editing.
\section*{Data Availability Statement}
The data that support the findings of this study are available from the corresponding author upon reasonable request.
\nocite{*}
\bibliographystyle{unsrt}
\bibliography{arxiv}  

\end{document}

%% file: sections/section_1.tex
\section{\label{sec1}Introduction}
To achieve the goal of global carbon neutrality, the vigorous development of wind energy has become a global consensus \cite{1veers2019grand}. When wind blows past a wind turbine, a wake region with reduced wind speed and increased turbulence is formed downstream of the turbine \cite{2stevens2017flow}. The wind speed deficit leads to a decrease in the inflow wind speed to downstream wind turbines, thereby resulting in power loss. The enhanced turbulence makes downstream wind turbines more susceptible to exposure to high-turbulence inflow, thereby increasing the fatigue load and potentially reducing their lifespan \cite{3meneveau2019big}. Since wind turbines operate in the surface layer of the atmospheric boundary layer (ABL), the evolution of wind-turbine wakes is significantly influenced by wind shear, wind veer, and atmospheric stability. Consequently, the wakes exhibit a highly complex, anisotropy three-dimensional structure, which poses a major challenge to accurately predicting the aforementioned physical quantities within wind-turbine wakes \cite{4porte2020wind}.

Since wind speed is directly related to the power generation of wind turbines, the velocity deficit in the wake region of wind turbines has received considerable attention from both wind energy academia and industry \cite{4porte2020wind}. Now, A relatively clear understanding of the spatial distribution of the velocity deficit has been achieved, and the relevant prediction models are also relatively complete, which are applicable to various complex situations. The derivation of the wake velocity deficit model generally depends on the theory of mass conservation or momentum conservation within the control volume, as well as the self-similarity of the velocity deficit profile. The self-similarity implies that the spatial distribution of the velocity deficit in the wake region satisfies a certain functional form. By assuming different profile forms and applying different physical conservation theories during the derivation process, numerous wake velocity deficit models have been proposed. For instance, Jensen\cite{5jensen1983note} and Frandsen et al.\cite{6frandsen2006analytical} both assumed that the profile of the wake velocity deficit satisfies the top-hat distribution and derived the Jensen model and the Frandsen model based on mass conservation and momentum conservation, respectively. As the understanding of the spatial distribution of the wake velocity deficit has deepened, researchers have gradually found that the top-hat distribution does not align with physical reality. Numerous wind tunnel experiments and numerical simulations show that the far wake velocity deficit exhibits the characteristics of a self-similar Gaussian distribution. Based on this, Bastankhah \& Port{\'e}-Agel\cite{7bastankhah2014new} assumed that the full wake satisfies the Gaussian distribution and derived the widely used Gaussian model based on momentum conservation theory. With further in-depth investigation of the near-wake of wind turbines, researchers have found that the near-wake of wind turbines does not satisfy the Gaussian distribution. Therefore, Keane\cite{8keane2021advancement}, Schreiber et al.\cite{9schreiber2020brief}, and Li et al.\cite{10li2025pressure} all assumed that the velocity deficit follows the double Gaussian distribution in the near-wake region and then slowly transitions to the Gaussian distribution in the far-wake region, leading to the development of the double Gaussian model. Meanwhile, Shapiro et al.\cite{11shapiro2019wake}, Blondel et al.\cite{12blondel2020alternative}, and Vahidi \& Port{\'e}-Agel\cite{13vahidi2022physics} assumed that the velocity deficit satisfies the Super-Gaussian distribution in the near-wake region and then slowly transitions to the Gaussian distribution in the far-wake region, resulting in the development of the Super-Gaussian model. Because these two types of models have many free parameters, parameter calibration is required before application. Therefore, they have not yet been widely used in engineering applications. In general, through nearly five decades of development, the prediction models for the wake velocity deficit have now reached a relatively mature stage, but determining the free parameters (e.g., wake width, near-wake length) via a physics-based approach remains an open research question\cite{13vahidi2022physics,14cheng2018simple,15du2022physical}.

\begin{figure}
\centering
\includegraphics[width=1.0\textwidth]{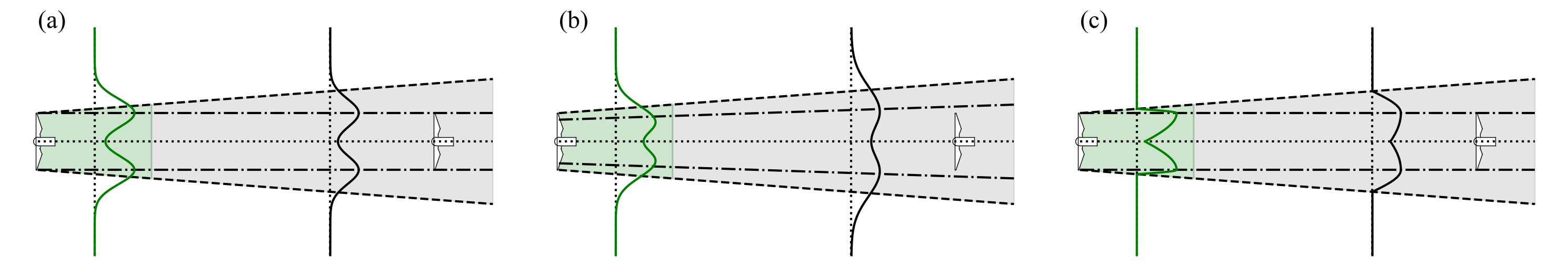}
\caption{Schematics of the assumed profiles of wake-added streamwise turbulence intensity for (a) the 3D-Gaussian model\cite{17ishihara2018new}, (b) the 3D-Gaussian-Cos model\cite{18li2022novel}, and (c) the 3D-Cos model\cite{19tian2022new}. (The dashed line represents the wake boundary, and the dot-dashed line indicates the peak position of wake-added turbulence.)}\label{fig1}
\end{figure}

Compared to the wake velocity deficit, the wind energy community has an insufficient understanding of the evolution and spatial distribution of wake-added turbulence. The prediction models for wake-added turbulence still lack a solid theoretical foundation. Most of these models are empirical and model only the added streamwise turbulence intensity component $\Delta I_u$. These empirical models typically begin by assuming the profile form of $\Delta I_u$, as shown in Figure \ref{fig1}, and then derive the relationship between the peak value or characteristic width of $\Delta I_u$ and inflow variables, wind turbine operating conditions and relative position through curve fitting based on wind tunnel experiments and numerical simulations. Earlier models of wake-added streamiwse turbulence intensity are all one-dimensional, that is, they only account for the streamwise variation. Among them, the model proposed by Crespo \& Hernandez\cite{16crespo1996turbulence} is the most classical. They assume that the peak value of $\Delta I_u$ is related to the wind turbine thrust coefficient, ambient turbulence intensity and streamwise location, and determine the tuning parameters in the model based on RANS simulation results. However, numerous wind tunnel experiments and numerical simulations show that wake-added streamwise turbulence intensity at hub height exhibits a dual-peak distribution \cite{17ishihara2018new,18li2022novel,19tian2022new}, which cannot be accurately represented by the conventional one-dimensional model. To address this, Ishihara et al. \cite{17ishihara2018new} assumed that wake-added streamwise turbulence intensity follows a double Gaussian distribution at hub height, with the peak value occurring downstream of the blade tip, and incorporated the ground effect in the vertical direction via a correction term, proposing the first three-dimensional wake-added streamwise turbulence intensity model (3D-Gaussian). The model's free parameters were obtained by fitting the results of wind tunnel experiments and numerical simulations. Based on the high-fidelity large-eddy simulation (LES) results, Li et al. \cite{18li2022novel} verified the dual-peak distribution characteristics of $\Delta I_u$ and found that its peak position expands linearly along the half-width of the wake, with the outer portion of the peak following the Gaussian distribution and the inner portion approximately following the Cosine function. Based on the above physical insights and further considering the ground effect, an improved three-dimensional wake-added streamwise turbulence intensity model (3D-Gaussian-Cos) was proposed. Recently, Tian et al. \cite{19tian2022new} also assumed that the peak of $\Delta I_u$ occurred downstream of the blade tip, but the spanwise influence extent extended further downstream. They also assumed that $\Delta I_u$ had a bimodal Cosine distribution profile, that is, both the outer and inner parts of the peak followed the Cosine function. Subsequently, considering the ground effect correction function, other free parameters were determined through quantitative analysis, and a novel three-dimensional wake-added streamwise turbulence intensity model (3D-Cos) was developed. These models have been widely used in engineering applications, such as the fast calculation of wind-farm flows \cite{20du2024momentum}, wind farm layout optimization \cite{21cao2022wind}, and wind turbine fatigue load assessment \cite{22peng2025layout}. However, it is worth noting that these empirical models rely on the assumption of a specific profile, with tuning parameters obtained through curve fitting. Therefore, the prediction accuracy of these models depends on the given profile, and their applicability is limited to the cases used for parameter fitting, demonstrating significant uncertainty in model predictions under unseen conditions. In addition to the models mentioned above, Jézéquel et al. \cite{23jezequel2024breakdown1,24jezequel2024breakdown2} used statistical wake meandering methods to develop a method for evaluating wake-added streamwise turbulence intensity. However, their model relies on physical quantities, such as wake width, wake center distribution width, and corrected mixing length, which are difficult to determine in practical applications.

Although streamwise turbulence intensity $I_u$ has many engineering applications, related studies \cite{25larsen2008wake,26feng2022componentwise,27noel2025wind,28li2024impacts} have shown that spanwise and vertical turbulence intensities ($I_v$ and $I_w$) are crucial for explaining the large-scale meandering phenomenon of the far wake and for theoretically determining wake width. Therefore, investigating the spatial distribution and streamwise evolution of turbulence kinetic energy (TKE) in the wake region holds significant theoretical and engineering value, and has only recently attracted attention in wind energy research. Similar to the modeling approach for $\Delta I_u$, Khanjari et al. \cite{29khanjari2025analytical} assumed that the normalized wake-added TKE in the wake region satisfies $\frac{k_w}{U_0^2}=\alpha\times A(x)\times G(r)\times W(z)$, and assumed that $A(x)$ and $W(z)$ approximately follow the Weibull-like distribution, while $G(r)$ follows the Gaussian distribution. The tuning parameters were determined based on numerous LES datasets, and finally a three-dimensional wake-added TKE model was proposed. However, this empirical method lacks a solid physical foundation and has limited physical interpretability. In addition, its applicability may be restricted to the working conditions of the fitted dataset, limiting its generalizability. To propose a physical-based wake-added TKE model, Bastankhah et al. \cite{30bastankhah2024modelling} studied the simplified wake-added TKE budget in cylindrical coordinates. They employed the gradient-diffusion and Boussinesq hypotheses to model the turbulent transport and shear production terms, respectively. Finally, they employed the Green's function method to derive the analytical solution of the modelled wake-added TKE budget and proposed an model in integral form of wake-added TKE. Validation based on wind tunnel experiments shows that the model can accurately predict the spatial distribution of wake-added TKE at hub height, provided that the wake velocity radial gradient, turbulent viscosity coefficient and TKE dissipation rate are directly measured. Therefore, two issues remain with this model: (1) The model is not closed, and nearly all model parameters are undetermined, preventing its direct application in engineering; (2) The model is applicable only at hub height and cannot accurately predict the three-dimensional spatial distribution of wake-added TKE induced by the ground. To address the second issue, Blondel et al. \cite{31blondel2025physics} further studied the wake-added TKE budget in Cartesian coordinates and, combining it with the self-similarity of wake-added TKE, derived a simple algebraic relation for calculating the three-dimensional distribution of wake-added TKE. The relevant modeling concept was further extended to the modeling of wake-added streamwise turbulence intensity, enabling physics-based modeling of wake-added turbulence. However, the derivation of the analytical model is limited by the assumption of self-similarity in wake-added TKE, which is only satisfied in the very far wake region $(x/D\ge7)$. Thus, their model exhibits significant uncertainty and demonstrates large prediction errors in the near-wake region. Furthermore, their model also suffers from the first issue, namely, the determination method for some key model parameters remained unspecified, preventing direct application in engineering. To address the aforementioned issues, we propose to decompose wake-added TKE into the azimuthally-averaged wake-added TKE part and a ground effect correction part, modeling these two parts separately. The calculation model for the azimuthally-averaged wake-added TKE is derived from the simplified azimuthally-averaged wake-added TKE budget, while the ground effect correction function is constructed by analogy with the ground effect correction function for wake-added streamwise turbulence intensity proposed by Li et al. \cite{18li2022novel}. Furthermore, we provide determination methods for all the free parameters, enabling the end-to-end prediction of the three-dimensional wake-added TKE provided that the basic inflow information and wind turbine operating information are given, significantly enhancing the engineering application value of the proposed model.

The remainder of the paper is structured as follows. In section \ref{sec2}, the spatial distribution of wake-added TKE based on LES results is investigated. In section \ref{sec3}, the simplified azimuthally-averaged wake-added TKE budget is derived, followed by the modelling of the azimuthally-averaged wake-added TKE. The modeling method for the ground effect correction function is presented in section \ref{sec4}. A comprehensive model validation is presented in section \ref{sec5} based on the LES data from this study and publicly available datasets from the literature. Finally, the concluding remarks are presented in section \ref{sec6}.

%% file: sections/section_2.tex
\section{Analysis of the spatial distribution of wake-added turbulence kinetic energy based on LES results}\label{sec2}

In this section, we investigate the spatial distribution and streamwise evolution of wake-added TKE based on LES results, and propose a fundamental decomposition approach for modeling wake-added TKE.

\subsection{Large-eddy simulation methodology}\label{sec21}

The LES solver used in this paper is a modified version of the open-source pseudospectral solver LESGO, which has been widely applied in the simulation of ABL flows \cite{32liu2021geostrophic,33narasimhan2024analytical}, wind-turbine and wind-farm flows \cite{34zhu2025jhtdb,35calaf2010large}, and urban street flows \cite{36fan2021impacts,37ge2020study}. Its simulation accuracy in wind-turbine wakes has been validated through wind tunnel experiments \cite{38stevens2018comparison} and compared with similar codes \cite{39martinez2018comparison}. To simplify the derivation of the wake-added TKE model, we neglect the effects of the Coriolis force and thermal stratification, and simulate the evolution of wind-turbine wakes under the pressure-driven neutral boundary layer (PDBL). The governing equations solved by the solver are based on the Cartesian coordinate system $xyz$, as follows:

\begin{eqnarray}
  \frac{\partial \tilde{u}_i}{\partial x_i} &=& 0 \\
  \frac{\partial \tilde{u}_i}{\partial t}+\tilde{u}_j\frac{\partial \tilde{u}_i}{\partial x_j} &=& -\frac{\partial \tilde{p}^*}{\partial x_i}-\frac{\partial \tau_{ij}^d}{\partial x_j}+\frac{f_i}{\rho}+F_p\delta_{i1}
\end{eqnarray}

where $\tilde{\cdot}$ represents the spatial filter at the grid scale $\Delta$, $t$ is time, $\tilde{u}_i$ is the instantaneous velocity in the $i$ direction (with $i=1,2,3$ corresponding to the streamwise ($x$), spanwise ($y$), and vertical ($z$) directions, respectively. $(\tilde{u}_1,\tilde{u}_2,\tilde{u}_3)$ and $(\tilde{u},\tilde{v},\tilde{w})$ are used interchangeably. $\tilde{p}^*=\tilde{p}/\rho+1/3\tau_{kk}$ is the modified pressure, where $\rho$ is the air density, and $\tau_{ij}$ is the subgrid stress. The deviatoric part of the subgrid stress, $\tau_{ij}^d$, is y the Smagorinsky model with wall damping function \cite{40bou2005scale}. $f_i$ is the body force induced by the wind turbine, $\delta_{ij}$ is the Kronecker symbol, and $F_p$ is the imposed pressure gradient. The solver employs a pseudospectral method in the horizontal directions, while the vertical derivatives are approximated with second-order central differences. For time advancement, a second-order-accurate Adams-Bashforth scheme is adopted. For clarity, the filter symbol is omitted hereafter.

\subsection{Case setups}\label{sec22}

To investigate the spatial evolution of wake-added TKE, we set up simulations of wind-turbine wakes under varying surface roughness ($z_0$). The simulation cases were based on the neutral boundary layer conditions of Abkar et al. \cite{41abkar2015influence} and Du et al. \cite{42du2021influence}, with an ABL height $\delta$ set to 500 m. The wind turbine is modeled using the filtered actuator disk model (ADM) proposed by Shapiro et al. \cite{43shapiro2019filtered}. The rotor diameter $D$ of the wind turbine is 100 m, and the hub height $H$ is also 100 m. The thrust coefficient $C_T$ of the wind turbine was set to 0.75, and the corresponding local thrust coefficient $C_{T}^\prime$ was approximately 1.33.

The simulation of the stand-alone wind turbine was performed using the concurrent precursor method, where the computational domain consists of both the precursor domain and the main domain. The precursor domain generates fully-developed turbulent inflow, while the main domain simulates the wind turbine. Both domains used the same grid settings and boundary conditions. Consistent with the horizontal pseudospectral discretization, the horizontal boundaries adopt periodic boundary conditions. The upper boundary condition is a stress-free condition, while the bottom surface adopts the wall stress model based on the Monin-Obukhov similarity theory \cite{44moeng1984large}. To minimize the influence of the top boundary on the simulation results, we set a buffer layer in the top 500m region following the methods of Bempedelis et al. \cite{45bempedelis2023turbulent} and Zhu et al. \cite{46zhu2025analytical}. The pressure gradient $F_p=u_*^2/\delta$ was applied only within the ABL height $\delta$, ensuring an average inflow wind speed $U_0$ of 8m/s at hub height. The turbulence generated in the precursor domain served as the inflow condition and was synchronized with the main domain in real-time via the fringe region, thereby avoiding the downwind flow affecting the flow upwind of the wind turbine due to periodic boundary conditions. This is a standard approach for simulating stand-alone wind-turbine wakes using a pseudospectral solver \cite{47stevens2014concurrent,48lanzilao2023improved}. The computational domain size was set to $3.2 \times 1.6 \times 1.0$ km$^3$, with grid points of $200 \times 160 \times 192$, uniformly distributed in all three directions were. This grid resolution ensured that more than 10 grids were included per rotor diameter, meeting the minimum grid resolution requirements for the simulation of wind-turbine wakes based on the ADM \cite{49wu2011large}. The key parameters of the LES cases are summarized in table \ref{tab1}.

\begin{table}
\centering
\renewcommand{\arraystretch}{1.25}
\caption{Overview of the LES case setups}\label{tab1}
\begin{tabular*}{0.80\linewidth}{@{}lccccc@{}}
\toprule
Calibration Case & \makecell[c]{Computation domain size\\ $L_x\times L_y\times L_z$ (km)} & \makecell[c]{Grid points\\ $N_x\times N_y\times N_z$} & $z_0$ (m) & $I_u$ &
$TI$ \\
\midrule
NBL-1 & & & 0.1 & 0.089 & 0.071 \\
NBL-2 & & & 0.05 & 0.081 & 0.064 \\
NBL-3 & \(3.2\times 1.6\times 1.0\) & \(200\times160\times192\) & 0.001
& 0.061 & 0.047 \\
NBL-4 & & & 0.0001 & 0.054 & 0.041 \\
\bottomrule
\end{tabular*}
\end{table}

\subsection{Spatial distribution of wake-added turbulence kinetic energy}\label{sec23}

\begin{figure}
\centering
\includegraphics[width=0.8\textwidth]{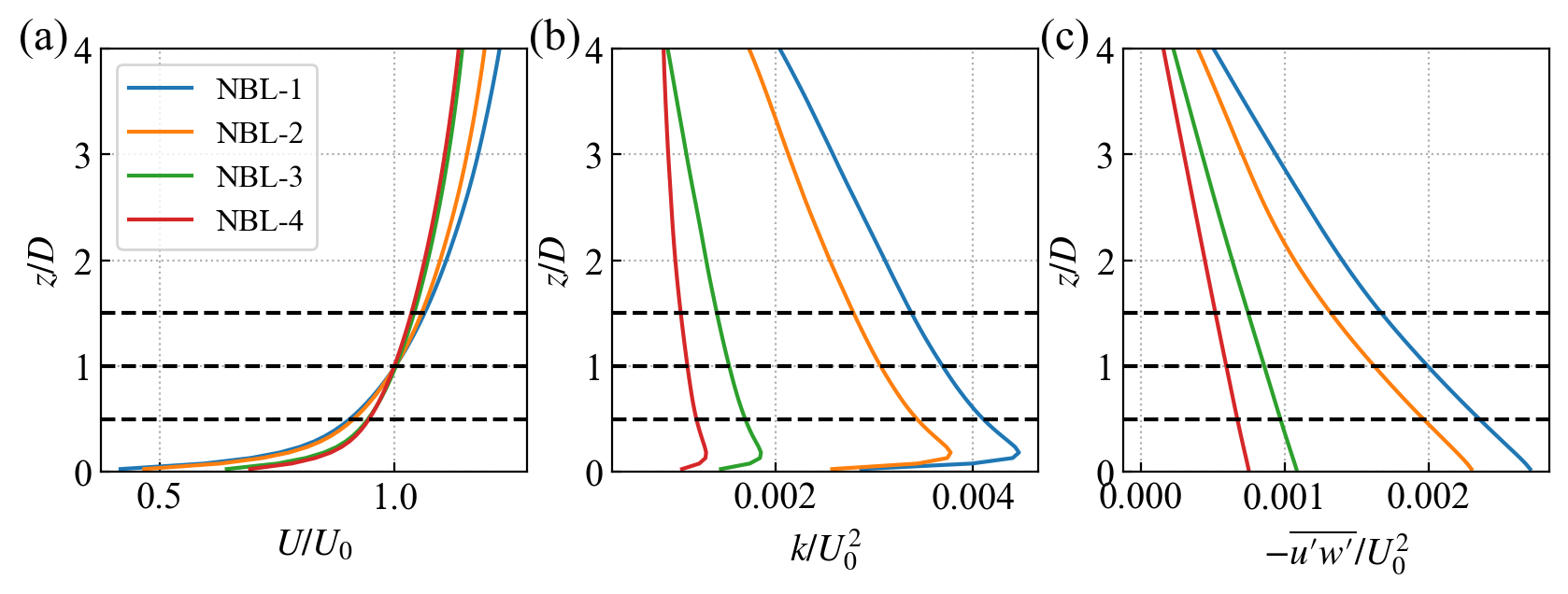}
\caption{Vertical profiles of the time- and horizontally-averaged (a) normalized inflow wind velocity, (b) normalized TKE, and (c) normalized shear stress in the precursor domain for different cases. The horizontal dashed lines indicate the top-tip, hub, and bottom-tip heights.}\label{fig2}
\end{figure}

Figure \ref{fig2} shows the inflow profiles for the LES cases considered in this study. It can be observed that we successfully simulated ABLs with varying inflow turbulence and shear stress. As excepted, the normalized TKE increases as the surface roughness $z_0$ increases. Within the ABL height, the shear stress decreases quasi-linearly to zero. Table \ref{tab1} presents the total turbulence intensity, $TI =\frac{\sqrt{2/3k}}{U_0} =\frac{\sqrt{(\overline{u^\prime u^\prime}+\overline{v^\prime v^\prime}+\overline{w^\prime w^\prime})/3}}{U_0}$, and the streamwise turbulence intensity $I_u =\frac{\sqrt{\overline{u^\prime u^\prime}}}{U_0}$, at hub height for different cases, which serves as key inputs for the proposed wake-added TKE model. Based on our LES results, we assume that  $I_u\approx1.28TI$ under neutral atmospheric conditions. This empirical relation is also supported by the LES results of Vahidi \& Port{\'e}-Agel \cite{13vahidi2022physics,50vahidi2025influence}. This empirical relation will be crucial for determining the input parameters of the proposed model, as most physical parameters (e.g., wake width and near-wake length) in the wake velocity deficit model are currently determined based on the streamwise turbulence intensity.

\begin{figure}
\centering
\includegraphics[width=0.9\textwidth]{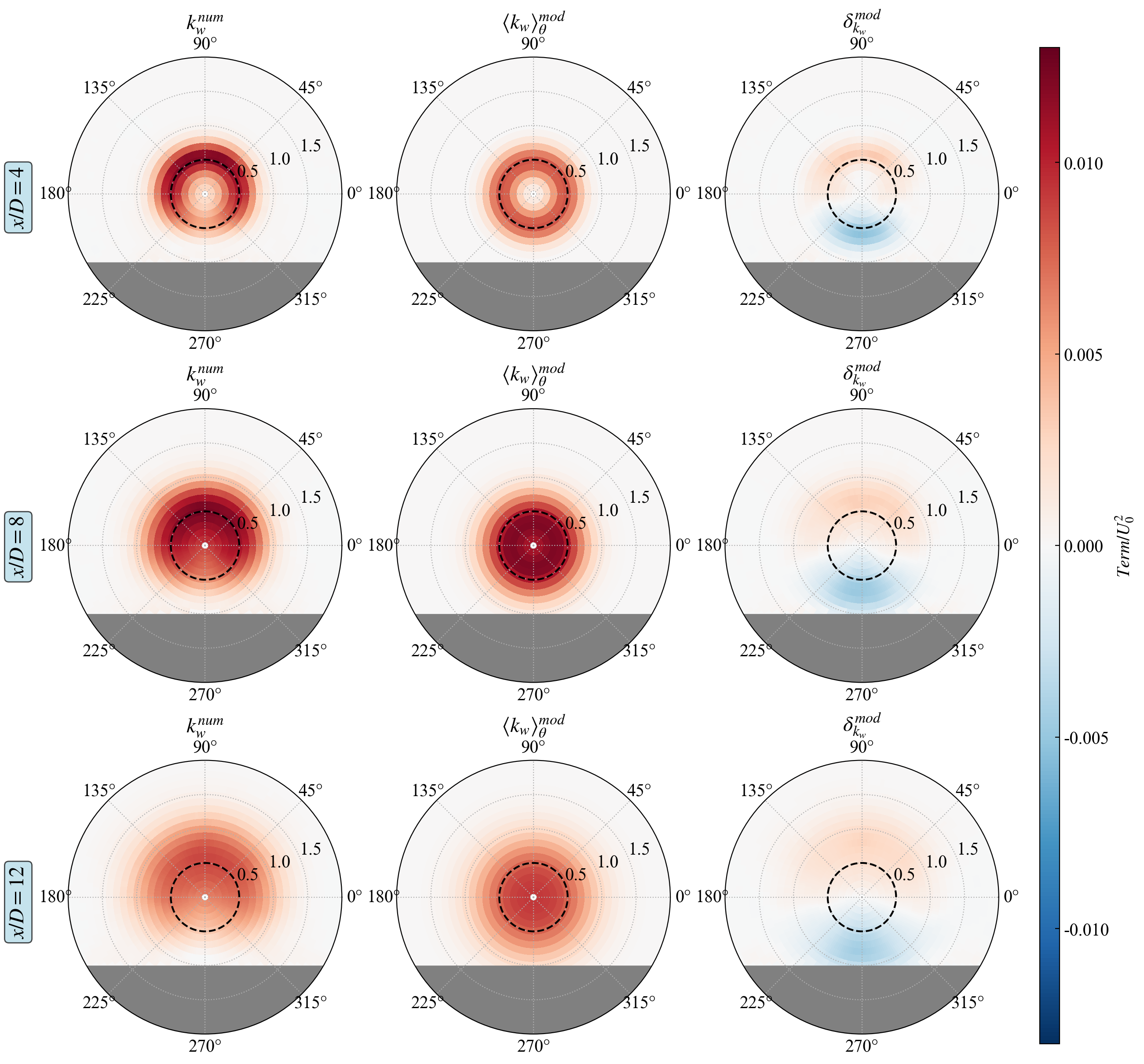}
\caption{Contourfs of (a1-a3) the wake-added TKE, (b1-b3) the azimuthally-averaged wake-added TKE, and (c1-c3) the ground effect correction term at different locations downstream of the wind turbine for the NBL-3 case. The black dashed line represents the boundary of the rotor swept region.}\label{fig3}
\end{figure}

Although the LES results are obtained using Cartesian coordinates $(x,y,z)$, we can easily transform the simulated variables into cylindrical coordinates $(x,r,\theta)$, where $x, r$ and $\theta$ correspond to the streamwise, radial, and azimuthal directions, respectively. Here, the origin of the cylindrical coordinate system is located at the center of the rotor, which simplifies the development of a physics-based azimuthally-averaged wake-added TKE model. The instantaneous velocity components in cylindrical coordinates are $(u_x, u_r, u_\theta)$, and the corresponding velocity fluctuation components are $(u_x^\prime, u_r^\prime, u_\theta ^\prime)$. The TKE is defined as $k=\frac{1}{2}(\overline{u_x^\prime u_x^\prime}+\overline{u_r^\prime u_r^\prime}+\overline{u_\theta^\prime u_\theta^\prime})$, and the wake-added TKE $k_w$ is defined as:

\begin{equation}\label{eq3}
  k_w = k - k^B
\end{equation}

where $k$ and $k^B$ represent the TKE in the main domain and the precursor domain, respectively. Figure \ref{fig3}(a1-a3) shows the contourfs of the wake-added TKE at different locations downstream of the wind turbine for the NBL-3 case. It can be seen that the wake-added TKE is approximately symmetrically distributed about the $y$-axis, but asymmetrically distributed about the hub height plane $z=H$, with the magnitude of the wake-added TKE in the upper sector significantly higher than that in the lower sector. This is mainly due to the asymmetry of the TKE generation term caused by the vertical shear of the inflow profiles in the ABL, as shown in figure \ref{fig2}(a). This indicates that the wake-added TKE exhibits three-dimensional anisotropic characteristics. Therefore, developing a three-dimensional wake-added TKE model is essential to improving the prediction accuracy of TKE in the wake region.

Mathematically, the spatial distribution of the three-dimensional wake-added TKE $k_w(x,r,\theta)$ can be viewed as an azimuthally-averaged wake-added TKE $\langle k_w\rangle_{\theta}(x,r)=\frac{1}{2\pi}\int_0^{2\pi} k_w(x,r,\theta)d\theta$ superimposed with a ground effect correction term $\delta_{k_w}(x,r,\theta)$. The spatial distribution of the azimuthally-averaged wake-added TKE and the ground effect correction term for the NBL-3 case are shown in figure \ref{fig3}. From figure \ref{fig3} (c1-c3), it can be seen that the ground effect correction term is positive above the hub height and negative below the hub height. Moreover, the peak value in the upper half is slightly smaller than that in the lower half, and the extent of influence in the upper half is significantly larger than in the lower half. Additionally, the ground effect correction term approximately follows the Gaussian distribution in the radial direction and the sinusoidal distribution in the azimuthal direction, which is consistent with the findings of Li et al. \cite{18li2022novel} regarding the ground correction function for wake-added streamwise turbulence intensity. This behavior is expected, as the dominant term in the wake-added TKE is $\Delta \overline{u_x^\prime u_x^\prime}$. Therefore, the spatial distribution of the wake-added TKE and streamwise turbulence intensity share some similarities, as confirmed by Blondel et al. \cite{31blondel2025physics}. This similarity provides a convenient basis for developing a model for the TKE ground effect correction term based on the existing model for its streamwise component. The spatial distribution of the wake-added TKE and the ground effect correction term for other cases is very similar to this case and is therefore not shown here.

Therefore, in the modeling of the wake-added TKE, it is necessary to construct submodules for $\langle k_w\rangle_{\theta}(x,r)$ and $\delta_{k_w}(x,r,\theta)$ separately. The $\langle k_w\rangle_{\theta}(x,r)$ submodule can be physically determined through a simplified azimuthally-averaged wake-added TKE budget, while the determination of $\delta_{k_w}(x,r,\theta)$ depends on the assumed spatial distribution function and empirical parameters. The following two sections will introduce these two submodules in detail. 

%% file: sections/section_3.tex
\section{Modeling of the azimuthally-averaged wake-added turbulence kinetic energy}\label{sec3}

To provide a method for calculating the azimuthally-averaged wake-added TKE $\langle k_w\rangle_{\theta}(x,r)$, we resorted to the TKE budget in cylindrical coordinates and derived the simplified azimuthally-averaged wake-added TKE budget. Based on this, we obtained the analytical solution for $\langle k_w\rangle_{\theta}(x,r)$ following Bastankhah et al. \cite{30bastankhah2024modelling}. Finally, we provided a method for determining the free parameters used in calculating $\langle k_w\rangle_{\theta}(x,r)$.

\subsection{Azimuthally-averaged wake-added turbulence kinetic energy budget}\label{sec31}

Similar to the momentum equation for the stand-alone wind-turbine wakes, the TKE budget serves as the basic governing equation for gaining physical insight into the evolution of wake-added TKE, which has been applied in the modeling of wake-added TKE \cite{30bastankhah2024modelling,31blondel2025physics}. In Bastankhah et al.\cite{30bastankhah2024modelling}, they assumed that the wind-turbine wake was axisymmetric. Based on the simplified wake-added TKE budget in cylindrical coordinates, they obtained an analytical expression of the wake-added TKE. This expression can accurately predict the spatial distribution of wake-added TKE at hub height (i.e., $\theta= 0^\circ$ and $180^\circ$). However, utility-scale wind turbines operate in the surface layer of the ABL, where they experience significant vertical wind shear. Therefore, the spatial distribution of the wake-added TKE exhibits strong three-dimensional anisotropic characteristics and does not satisfy the basic assumption of the axisymmetric wake. Therefore, their model cannot be directly applied in the ABL and require improvement. To start with, we will investigate the TKE budget in cylindrical coordinates using the high-fidelity LES results of the utility-scale wind turbine, thereby deriving the azimuthally-averaged wake-added TKE budget. The resolved-scale TKE budget in cylindrical coordinates based on the LES results is written as:

\begin{equation}\label{eq4}
\begin{aligned}
&\underbrace{\bar{u}_x\frac{\partial k}{\partial x}+\bar{u}_r\frac{\partial k}{\partial r}+\frac{\bar{u}_\theta}{r}\frac{\partial k}{\partial \theta}}_{\mathcal{A}}+\underbrace{\frac{\partial \overline{k^\prime u_x^\prime}}{\partial x}+\frac{1}{r}\frac{\partial (r\overline{k^\prime u_r^\prime})}{\partial r}+\frac{1}{r}\frac{\partial \overline{k^\prime u_\theta^\prime}}{\partial \theta}}_{\mathcal{T}_t}+\underbrace{\frac{\partial \overline{p^{*\prime} u_x^\prime}}{\partial x}+\frac{1}{r}\frac{\partial (r\overline{p^{*\prime} u_r^\prime})}{\partial r}+\frac{1}{r}\frac{\partial \overline{p^{*\prime} u_\theta^\prime}}{\partial \theta}}_{\mathcal{T}_p}+\mathcal{T}_{sgs}+{\mathcal{D}} \\
&\quad =\underbrace{\overline{u_r^\prime u_\theta^\prime}\frac{\bar{u}_\theta}{r}-\overline{u_\theta^\prime u_\theta^\prime}\frac{\bar{u}_r}{r}-\left( \overline{u_i^\prime u_x^\prime}\frac{\partial}{\partial x}+ \overline{u_i^\prime u_r^\prime}\frac{\partial}{\partial r}+\frac{\overline{u_i^\prime u_\theta^\prime}}{r}\frac{\partial }{\partial \theta}\right)\bar{u}_i}_{\mathcal{P}_s}+\underbrace{\frac{\overline{f_x^\prime u_x^\prime}}{\rho}+\frac{\overline{f_r^\prime u_r^\prime}}{\rho}+\frac{\overline{f_\theta^\prime u_\theta^\prime}}{\rho}}_{\mathcal{P}_t}
\end{aligned}
\end{equation}

where $ k^\prime=(u_x^\prime u_x^\prime+u_r^\prime u_r^\prime+u_\theta^\prime u_\theta^\prime)/2$ is the resolved-scale fluctuating TKE, and $f_x^\prime,f_r^\prime$ and $f_\theta^\prime$ represent the streamwise, radial and azimuthal body force fluctuations of the wind turbine, respectively. $\mathcal{A}$ is the advection by the mean flow, $\mathcal{T}_t$ is the transport by turbulent velocity fluctuations, $\mathcal{T}_p$ is the transport by pressure fluctuations, $\mathcal{T}_{sgs}$ is the transport by the subgrid-scale stress fluctuations, $\mathcal{D}$ is the resolved-scale TKE dissipation, $\mathcal{P}_s$ is the shear production (where $i=x,r,\theta,$ and Einstein's summation convection is applied), and $\mathcal{P}_t$ is the production induced by the body force fluctuations. In the above equation, the advection term $\mathcal{A}$, the turbulent transport term $\mathcal{T}_t$, the pressure transport term $\mathcal{T}_p$, the shear production term $\mathcal{P}_s$, and the wind turbine production term $\mathcal{P}_t$ can be indirectly calculated using the basic physical quantities in Cartesian coordinates. In studies of ABL flows and wind-turbine wakes, the subgrid-scale stress transport term $\mathcal{T}_{sgs}$ is typically negligible and can be ignored compared to other terms \cite{31blondel2025physics,51centurelli2025asymmetric,52klemmer2024momentum}. Furthermore, to mitigate the influence of numerical dissipation, the resolved-scale dissipation term $\mathcal{D}$ is generally calculated by other terms, namely, $\mathcal{D}=\mathcal{P}_s+\mathcal{P}_t-(\mathcal{A}+\mathcal{T}_t+\mathcal{T}_p+\mathcal{T}_{sgs})$, which is a common practice for calculating the dissipation term in LES results \cite{31blondel2025physics,51centurelli2025asymmetric}. Therefore, based on the LES results in Cartesian coordinates, we can derive the corresponding TKE budget in cylindrical coordinates, which will serve as the physical basis for understanding the spatial distribution of the three-dimensional wake-added TKE and proposing an analytical model. It should be noted that the above equation is not only applicable to the wind-turbine wakes but also to the atmospheric boundary layer. Similar to Blondel et al. \cite{31blondel2025physics} and Klemmer \& Howland \cite{52klemmer2024momentum}, we can define the wake-added TKE budget as:

\begin{equation}\label{eq5}
  \underbrace{\mathcal{A}-\mathcal{A}^B}_{\mathcal{A}^w}+\underbrace{\mathcal{T}_t-\mathcal{T}_t^B}_{\mathcal{T}_t^w}+\underbrace{\mathcal{T}_p-\mathcal{T}_p^B}_{\mathcal{T}_p^w}+\underbrace{\mathcal{D}-\mathcal{D}^B}_{\mathcal{D}^w}=\underbrace{\mathcal{P}_s-\mathcal{P}_s^B}_{\mathcal{P}_s^w}+\underbrace{\mathcal{P}_t-\mathcal{P}_t^B}_{\mathcal{P}_t^w}
\end{equation}
where the superscript $^B$ represents the term in the precursor domain, and the physical quantities in the term also correspond to the physical quantities in the precursor domain. Correspondingly, $\mathcal{A}^w,\mathcal{T}_t^w.\mathcal{T}_p^w,\mathcal{D}^w,\mathcal{P}_s^w$, and $\mathcal{P}_t^w$ can be viewed as the wake-added advection, turbulent transport, pressure transport, TKE dissipation, shear production and wind turbine production terms. By comparing and analyzing the wake-added TKE budget term by term (not shown here), we found that the influence region of the wake-added pressure transport term $\mathcal{T}_p^w$ and the wake-added wind turbine production term $\mathcal{P}_t^w$ is limited to the vicinity of the wind turbine. In contrast, the wake-added advection term $\mathcal{A}^w$, the wake-added turbulent transport term $\mathcal{T}_t^w$, the wake-added dissipation term $\mathcal{D}^w$, and the wake-added shear generation term $\mathcal{P}_s^w$ play a dominant role throughout wind-turbine wakes. This is consistent with the findings of Blondel et al. \cite{31blondel2025physics} and Klemmer \& Howland \cite{52klemmer2024momentum}. Therefore, we can obtain the following simplified wake-added TKE budget:
\begin{equation}\label{eq6}
  \mathcal{A}^w+\mathcal{T}_t^w+\mathcal{D}^w\approx\mathcal{P}_s^w
\end{equation}

\begin{figure}
\centering
\includegraphics[width=1.0\textwidth]{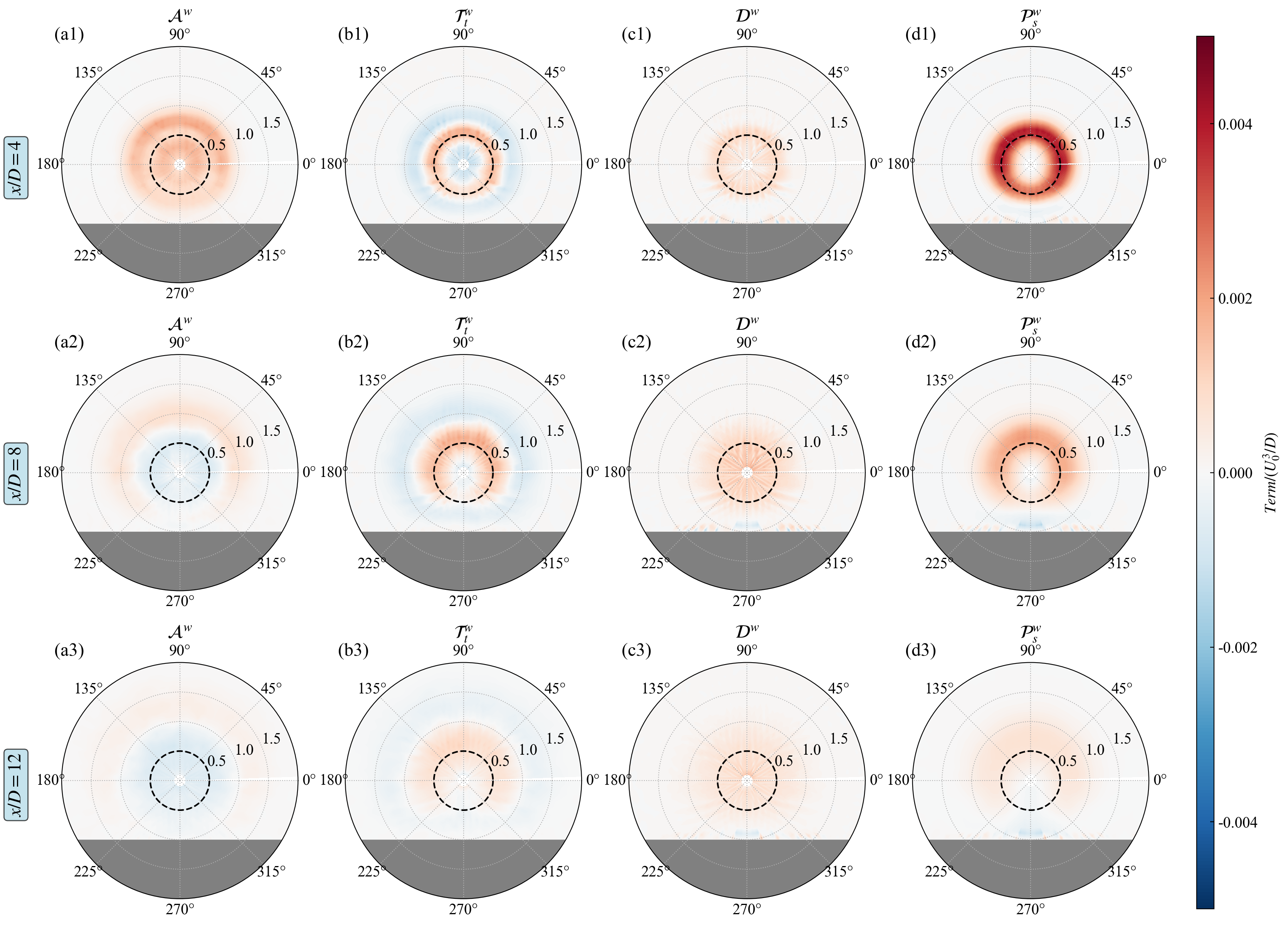}
\caption{Contourfs of (a1-a3) the wake-added advection term $\mathcal{A}^w$, (b1-b3) the wake-added turbulent transport term $\mathcal{T}_t^w$, (c1-c3) the wake-added dissipation term $\mathcal{D}^w$, and (d1-d3) the wake-added shear production term $\mathcal{P}_s^w$ of the simplified wake-added TKE budget at different locations downstream of the wind turbine for the NBL-3 case. The black dashed line represents the boundary of the rotor swept region.}\label{fig4}
\end{figure}

Figure \ref{fig4} shows the contourfs of different terms in the simplified wake-added TKE budget Eq. (\ref{eq6}) at different locations downstream of the wind turbine for the NBL-3 case. It can be seen that, due to vertical shear, the wake-added shear production term is higher in the upper sector than the lower sector, resulting in enhanced turbulent transport and advection. This explains the observation in figure \ref{fig3} (a1-a3) that the wake-added TKE in the upper sector is higher than that in the lower sector. In addition, we can see that $\mathcal{T}_t^w$ and $\mathcal{P}_s^w$ have a similar spatial distribution pattern at different downstream locations, with the peak magnitude decreasing and its influence region increasing at different downstream locations. The wake-added shear production has the highest value at the boundary of the rotor swept region, leading to the localized maximum value of wake-added TKE. As a result, $\mathcal{T}_t^w$ has the maximum value around the boundary of the rotor swept region, while $\mathcal{T}_t^w<0$ in other regions. In contrast, the sign of $\mathcal{A}^w$ varies at different downstream locations. At $x/D=4$, $\mathcal{A}^w>0$ in the entire region, indicating that the wake-added TKE has not reached its maximum value. At $x/D=8$ and $x/D=12$, $\mathcal{A}^w<0$ when $r/D<0.75$ and $r/D<1.00$, respectively, indicating that the wake-added TKE reaches its peak value between $x/D=4$ and $x/D=8$, and $k_w$ decreases streamwise in the region where $\mathcal{A}^w<0$. At the outer boundary of the wind-turbine wakes, $\mathcal{A}^w>0$ and is mainly balanced by the $\mathcal{T}_t^w$, with $k_w$ still increasing streamwise. In contrast, $\mathcal{D}^w>0$ throughout the considered region. The results of other cases are similar to this case, and therefore are not shown here. Overall, the simplified wake-added TKE budget successfully accounts for the processes of TKE production, transport, advection and dissipation in wind-turbine wakes.

However, directly solve the above simplified wake-added TKE budget for the three-dimensional wake-added TKE is still infeasible. Therefore, we will further derive the azimuthally-averaged wake-added TKE budget and provide the analytical expression for the azimuthally-averaged wake-added TKE based on this. After obtaining this variable, the three-dimensional spatial distribution of wake-added TKE can be easily obtained by introducing a ground effect correction term \cite{17ishihara2018new,18li2022novel,19tian2022new}. Here, we define the azimuthally-averaged operator $\langle\cdot\rangle_\theta$ as follows:

\begin{equation}\label{eq7}
  \langle Term\rangle_\theta(x,r)=\frac{1}{2\pi}\int_0^{2\pi}Term(x,r,\theta)d\theta
\end{equation}

Therefore, the simplified azimuthally-averaged wake-added TKE budget can be written as:
\begin{equation}\label{eq8}
  \langle\mathcal{A}^w\rangle_\theta+\langle\mathcal{T}_t^w\rangle_\theta+\langle\mathcal{D}^w\rangle_\theta\approx\langle\mathcal{P}_s^w\rangle_\theta
\end{equation}

Furthermore, as shown in Eq. (\ref{eq4}), the advection term $\mathcal{A}^w$, the turbulent transport term $\mathcal{T}_t^w$, and the shear production term $\mathcal{P}_s^w$ consist of multiple components. However, among these, one component is dominant. Therefore, the above equation can be further simplified as:

\begin{equation}\label{eq9}
  \underbrace{\left\langle\bar{u}_x\frac{\partial k}{\partial x}\right\rangle_\theta}_{\langle\mathcal{A}^w\rangle_\theta^d}+\underbrace{\left\langle\frac{1}{r}\frac{\partial (r\overline{k^{\prime} u_r^{\prime}})}{\partial r}\right\rangle_\theta}_{\langle\mathcal{T}_t^w\rangle_\theta^d}+\langle\mathcal{D}^w\rangle_\theta\approx\underbrace{-\left\langle\overline{u_x^{\prime}u_r^{\prime}}\frac{\partial \bar{u}_x}{\partial r}\right\rangle_\theta}_{\langle\mathcal{P}_s^w\rangle_\theta^d}
\end{equation}
where $\langle\mathcal{A}^w\rangle_\theta^d,\langle\mathcal{T}_t^w\rangle_\theta^d$, and $\langle\mathcal{P}_s^w\rangle_\theta^d$ represent the dominant components of the wake-added advection term $\mathcal{A}^w$, the wake-added turbulent transport term $\mathcal{T}_t^w$, and the wake-added shear production term $\mathcal{P}_s^w$, respectively.

\begin{figure}
\centering
\includegraphics[width=1.0\textwidth]{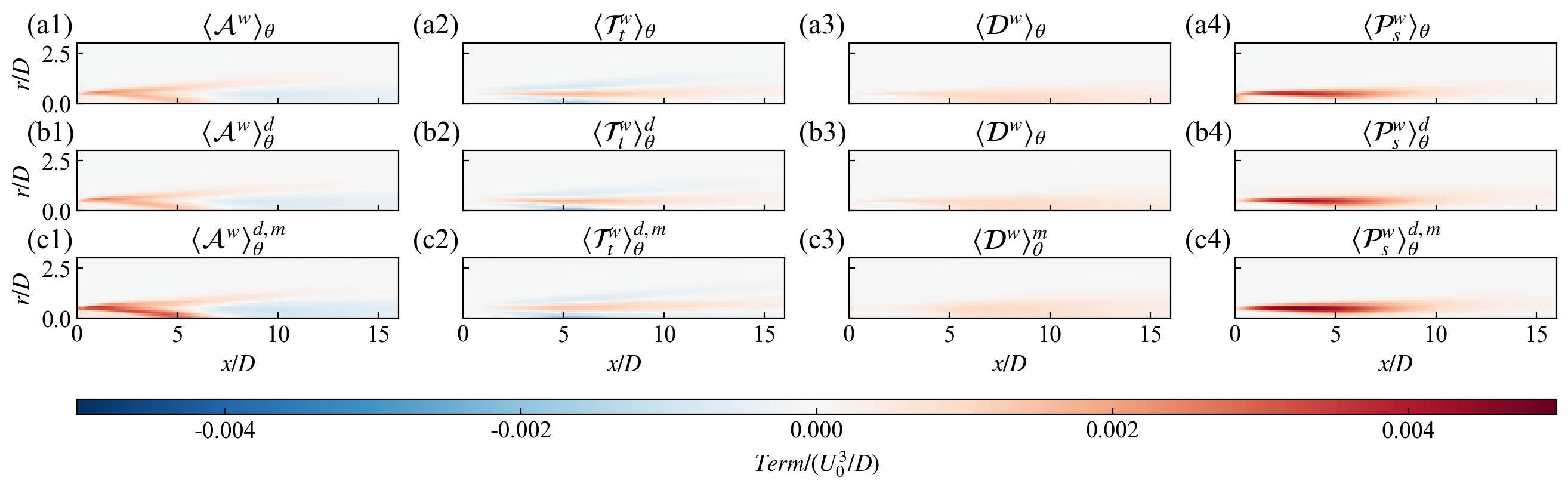}
\caption{Comparison of (a1-a4) the full terms in Eq. (\ref{eq6}), (b1-b4) their dominant components in Eq. (\ref{eq9}), and (c1-c4) the modeled dominant components in Eq. (\ref{eq13}) for the NBL-3 case.}\label{fig5}
\end{figure}

Figure \ref{fig5} shows the comparison of the full terms and their dominant components in the simplified azimuthally-averaged wake-added TKE budget for the NBL-3 case. It can be seen that the dominant components characterizes well the spatial distribution of the full terms, thereby validating the applicability of Eq. (\ref{eq9}). The results of other cases are similar to this case, so they are not shown here.

\subsection{Analytical modelling of the azimuthally-averaged wake-added turbulence kinetic energy budget}\label{sec32}

The simplified azimuthally-averaged wake-added TKE budget in Eq. (\ref{eq9}) still contains unresolved terms such as $\overline{k^\prime u_r^\prime}$, $\overline{u_x^\prime u_r^\prime}$, and $\langle\mathcal{D}^w\rangle_\theta$, which are difficult to solve directly. To solve the above budget analytically, it is essential to model these terms. According to Bastankhah et al. \cite{30bastankhah2024modelling} and Blondel et al. \cite{31blondel2025physics}, the wake-added turbulent transport term $\langle\mathcal{T}_t^w\rangle_\theta^d$ can be modeled based on the gradient-diffusion hypothesis, that is:

\begin{equation}\label{eq10}
  \underbrace{\left\langle\frac{1}{r}\frac{\partial (r\overline{k^\prime u_r^\prime})}{\partial r}\right \rangle_\theta}_{\langle\mathcal{T}_t^w\rangle_\theta^d}\approx-\frac{1}{r}\frac{\partial}{\partial r}\left(r\nu_t\frac{\partial \langle k_w\rangle_\theta(x,r)}{\partial r}\right)
\end{equation}
where $\nu_t=\nu_t(x)$ is the turbulent viscosity coefficient, which can be obtained by fitting the Reynolds shear stress and velocity gradient at hub height. The Reynolds shear stress $\overline{u_x^\prime u_r^\prime}$ can be modeled using the Boussinesq hypothesis, i.e.
\begin{equation}\label{eq11}
  \overline{u_x^\prime u_r^\prime}=-\nu_t\left(\frac{\partial \bar{u}_x}{\partial r}+\frac{\partial \bar{u}_r}{\partial x}\right)
\end{equation}
where $\frac{\partial \bar{u}_x}{\partial r}>>\frac{\partial \bar{u}_r}{\partial x}$. Therefore, we have
\begin{equation}\label{eq12}
  \underbrace{-\left\langle\overline{u_x^{\prime}u_r^{\prime}}\frac{\partial \bar{u}_x}{\partial r}\right\rangle_\theta}_{\langle\mathcal{P}_s^w\rangle_\theta^d}\approx \nu_t(x)\left\langle\left(\frac{\partial \bar{u}_x}{\partial r}\right)^2\right\rangle_\theta\approx \nu_t(x)\left(\underbrace{\frac{\partial \bar{u}_x(x,r,\theta=0^\circ)}{\partial r}}_{U_\rho}\right)^2
\end{equation}
Furthermore, the azimuthally-averaged square of radial velocity gradient can be simplified to the square of the radial velocity gradient $U_\rho$ at hub height. According to Bastankhah et al. \cite{30bastankhah2024modelling}, the modeled wake-added TKE dissipation term $\langle\mathcal{D}^w\rangle_\theta^m$ can be expressed as $C_\varepsilon \langle k_w\rangle_\theta^{3/2}/l_m$, where $l_m=l_m(x)$ and $C_\varepsilon=C_\varepsilon(x) $ are the mixing length and the normalized TKE dissipation rate, respectively. Since the turbulent viscosity coefficient $\nu_t$ can be written in the form $c\langle k_w\rangle_\theta^{1/2}l_m$, and thereby $\langle\mathcal{D}^w\rangle_\theta^m=\nu_t\langle k_w\rangle_\theta/\Psi(x)$, where $\Psi(x)=cl_m^2/C_\varepsilon$. Furthermore, the wake-added advection term $\langle\mathcal{A}^w\rangle_\theta^d$ needs to be linearized, i.e., $\bar{u}_x$ is replaced by the freestream inflow velocity $U_0$ at hub height. Finally, we can obtain the modeled azimuthally-averaged wake-added TKE budget, namely:

\begin{equation}\label{eq13}
  \underbrace{U_0\frac{\partial \langle k_w\rangle_\theta(x,r)}{\partial x}}_{\langle\mathcal{A}^w\rangle_\theta^{d,m}}-\underbrace{\frac{\nu_t(x)}{r}\frac{\partial}{\partial r}\left(r\frac{\partial \langle k_w\rangle_\theta(x,r)}{\partial r}\right)}_{\langle\mathcal{T}_t^w\rangle_\theta^{d,m}}+\underbrace{\frac{\nu_t(x)}{\Psi(x)}\langle k_w\rangle_\theta(x,r)}_{\langle\mathcal{D}^w\rangle_\theta^m}\approx \underbrace{\nu_t(x)U_\rho^2}_{\langle\mathcal{P}_s^w\rangle_\theta^{d,m}}
\end{equation}

where the superscript $^m$ represent the modeled wake-added TKE budget term. The square of the mixing length $l_m^2$ in $\Psi(x) = cl_m^2/C_\varepsilon$ can be obtained by fitting $-\overline{u_x^\prime u_r^\prime}$  with $|\frac{\partial U}{\partial r}|\frac{\partial U}{\partial r}$ at hub height. After obtaining the streamwise variation of the turbulent viscosity coefficient $\nu_t(x) $and the mixing length $l_m(x)$, the coefficient $c$ at different locations can be calculated as $\frac{\nu_t(x)}{\langle k_w\rangle_\theta^{1/2}l_m(x)}$. Similarly, the normalized TKE dissipation rate $C_\varepsilon$ can be calculated as $\frac{\langle \mathcal{D}^w\rangle_\theta^m l_m(x)}{\langle k_w\rangle_\theta^{3/2}}$. By definition, both $c$ and $C_\varepsilon$ are two-dimensional functions relating to streamwise and radial directions, which poses a great challenge in analytically solve Eq. (\ref{eq13}). Therefore, when calculating these two key coefficients, we replace $\langle k_w\rangle_\theta(x,r$) and $\langle\mathcal{D}^w\rangle_\theta^m(x,r)$ with the maximum value of the azimuthally-averaged wake-added TKE, $K_w(x)$, within the rotor and the rotor- and azimuthally-averaged wake-added TKE dissipation rate $\langle\mathcal{D}^w\rangle_\theta^{m,r}(x)$, respectively. The resulting $c(x)$ and $C_\varepsilon(x)$ will then be one-dimensional streamwise functions. Therefore, $\Psi(x)=cl_m^2/C_\varepsilon$ is also a one-dimensional function. Figures \ref{fig5}(c1-c4) further illustrate the contourfs of different modeled terms in Eq. (\ref{eq13}) for the NBL-3 case, in which the variables $U_0$, $\langle k_w\rangle_\theta$, $\nu_t(x)$, $\Psi(x)$, and $U_\rho$ are all derived from LES results. It can be seen that, except for small errors caused by the linearization, the modeling of other terms effectively captures both the magnitude and spatial distribution of the terms, indicating that the modeled wake-added TKE budget in Eq. (\ref{eq13}) can accurately capture the production, advection, transport, and dissipation processes relevant to the azimuthally-averaged wake-added TKE.Therefore, Eq. (\ref{eq13}) provides an important theoretical basis for solving the azimuthally-averaged wake-added TKE.

Mathmetically, Eq. (\ref{eq13}) is a second-order non-homogeneous linear differential equation. To solve this equation, we can directly construct a two-dimensional grid and use numerical discretization methods to obtain the spatial distribution of $\langle k_w\rangle_\theta(x,r)$. It is worth noting that Eq. (\ref{eq13}) is consistent with the wake-added TKE budget for axisymmetric wakes proposed by Bastankhah et al. \cite{30bastankhah2024modelling}. Therefore, Eq. (\ref{eq13}) can be solved using the analytical solution method proposed by Bastankhah et al. \cite{30bastankhah2024modelling} based on the Green's function. Assuming $\langle k_w\rangle_\theta(0,r)=0$, $\partial\langle k_w\rangle_\theta(x,0)/\partial r=0$ and $\langle k_w\rangle_\theta(x,\infty)\rightarrow 0 \text{ as }r\rightarrow\infty$, the azimuthally-averaged wake-added TKE at any position $(x,r)$ downstream of the wind turbine can be written in the form of the following integral expression, i.e.

\begin{equation}\label{eq14}
  \langle k_w\rangle_\theta(x,r)=
\begin{cases}
\int_{X=x_T}^{x-\delta}\int_{\rho=0}^{3D}\frac{\nu_t(X)}{U_0}e^{-\Psi(X,x)}\left\{\frac{e^{-\frac{(r-\rho)^2}{4\phi(X,x)}}}{\sqrt{4\pi\phi(X,x)}}\right\}f\left(\frac{r\rho}{\phi(X,x)}\right)U_\rho^2(X,\rho)\sqrt{\frac{\rho}{r}}d\rho dX \qquad r>0\\
\int_{X=x_T}^{x-\delta}\int_{\rho=0}^{3D}\frac{\nu_t(X)}{2U_0\phi(X,x)}\exp\left\{-\frac{\rho^2}{4\phi(X,x)}-\Psi(X,x)\right\}U_\rho^2(X,\rho)\rho d\rho dX\qquad r=0
\end{cases}
\end{equation}
where
\begin{equation}\label{eq15}
  f(z)\approx\begin{cases}
\sqrt{\pi z}\exp(-z/2)(1+\frac{z^2}{16}+\frac{z^4}{1024}),\quad 0\leq z\leq 4\\
1+\frac{1}{4z}+\frac{9}{32z^2}, \qquad\qquad\qquad\qquad\quad z>4
\end{cases},
\qquad\phi(X,x)=\frac{1}{U_0}\int_{\xi=X}^x\nu_t(\xi)d\xi,
\quad \psi(X,x)=\frac{1}{U_0}\int_{\xi=X}^x\frac{\nu_t(\xi)}{\Psi(\xi)}d\xi
\end{equation}

In the about equations, $\rho,z,X,\xi,$ $\phi(X,x)$, and $\psi(X,x)$ are all dummy variables, and $f(z)$ is an approximation of the modified Bessel function of the first kind. The above equation shows that the azimuthally-averaged wake-added TKE at any point $(x,r)$ downstream of the wind turbine is a double integral over the integration region $\Omega=\{(X,\rho)|x_T\le X\le x-\delta,0\le \rho\le 3D\}$, where $x_T$ can be regarded as the streamwise coordinate of the wind turbine. In the coordinate system specified in this paper, $x_T/D=0$. It should be noted that only a simplified form for numerical integration is given here. For more information about the Green's function and its theoretical derivation, please refer to Bastankhah et al. \cite{30bastankhah2024modelling}. Eq. (\ref{eq14}) gives the calculation model of the azimuthally-averaged wake-added TKE, avoiding the cumbersome discrete solution process and effectively saving computational cost. Therefore, we will use this integration method to calculate the azimuthally-averaged wake-added TKE in this paper.

\subsection{Determination of the key parameters}\label{sec33}

Before applying Eq. (\ref{eq14}) to engineering applications, it is necessary to provide the methods for determining the unknown parameters $U_\rho(x,r)$, $\nu_t(x)$, and $\Psi(x)$. The radial wake velocity gradient $U_\rho(x,r)$ can be directly calculated using the analytical wake model. However, there are no existing method to determine the turbulent viscosity coefficient $\nu_t(x)$ and $\Psi(x)$ in the wind-turbine wakes. To this end, we will investigate the streamwise variation of $\nu_t(x)$ and $\Psi(x)$ based on the LES cases and develop empirical expressions for them based on the ambient turbulence variable.

\subsubsection{Wake velocity radial gradient $U_\rho$}\label{sec331}

\begin{figure}
\centering
\includegraphics[width=1.0\textwidth]{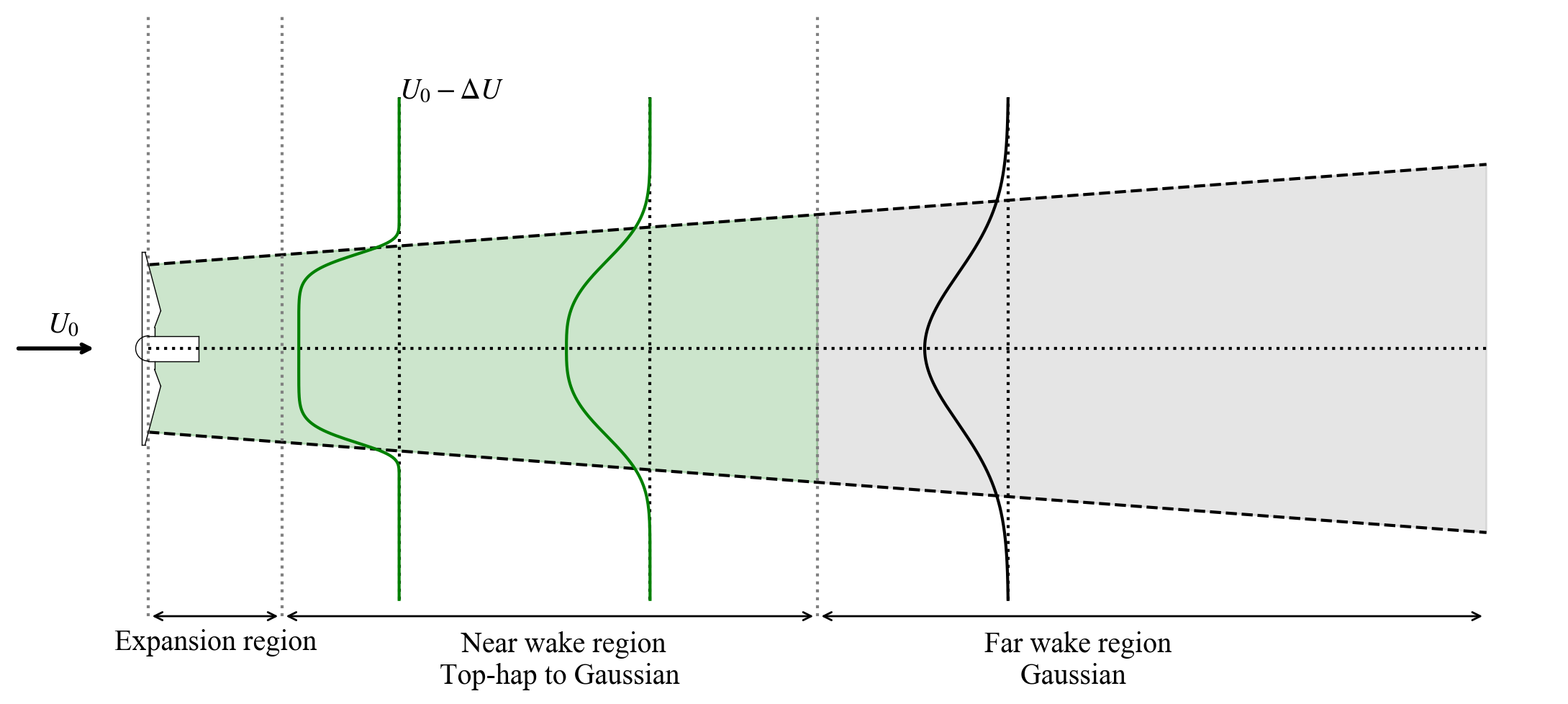}
\caption{Schematic of the Super-Gaussian wake model (The figure is modified from Vahidi \& Port{\'e}-Agel\cite{13vahidi2022physics}).}\label{fig6}
\end{figure}

The determination of the wake velocity radial gradient $U_\rho$ mainly depends on the accurate calculation of the wake velocity deficit. It is worth noting that conventional wake models only exhibit high prediction accuracy in the far wake region. As mentioned in Sec. \ref{sec32}, the azimuthally-averaged wake-added TKE depends on the double integral over $\Omega$. That is, the predicted wake-added TKE in the far-wake region also depends on the velocity gradient in the near-wake region. Therefore, the accurate modeling of the velocity deficit in the near-wake region is also crucial. Unlike the self-similar Gaussian profile in the far-wake region, the velocity deficit profile in the near-wake region generally follows a Super-Gaussian or double-Gaussian profile \cite{8keane2021advancement,9schreiber2020brief,10li2025pressure,11shapiro2019wake,12blondel2020alternative,13vahidi2022physics}. Here, we adopt the Super-Gaussian full wake model proposed by Vahidi \& Port{\'e}-Agel \cite{13vahidi2022physics}, as shown in figure \ref{fig6}.

Specifically, the Super-Gaussian wake model assumes that the wake velocity deficit profile gradually transitions from an approximately top-hat distribution to the Gaussian distribution at the onset of the far-wake region, i.e.,
\begin{equation}\label{eq16}
  \frac{\Delta U(x,r)}{U_0}=C(x)\exp\left(-\frac{r^{n(x)}}{2\sigma^2(x)}\right)
\end{equation}

where $\Delta U$ is the velocity deficit, $C(x)$ is the normalized maximum velocity deficit, $n(x)$ is the Super-Gaussian exponent, and $\sigma(x)$ represents the wake width. The onset of the far-wake region $x_{th}$, the corresponding wake width $\sigma_{th}$, and the normalized maximum velocity deficit $C_{th}$ at $x_{th}$ can be determined using the following expressions:
\begin{equation}\label{eq17}
  \frac{x_{th}}{D}=\frac{(1+\sqrt{1-C_T})}{\sqrt{2}(2.32I_u+0.154(1-\sqrt{1-C_T}))}\\
\qquad \frac{\sigma_{th}}{D}=k^*\frac{x_{th}}{D}+\varepsilon\\
\qquad C_{th}=1-\sqrt{1-\frac{C_T}{8(\sigma_{th}/D)^2}}
\end{equation}

where $k^*=0.01+0.28I_u$ and $\varepsilon=(0.1+0.1C_T)\sqrt{\frac{1+\sqrt{1-C_T}}{2\sqrt{1-C_T}}}$ are the wake growth rate and wake width intercept\cite{53blondel2023brief}, respectively. It should be noted that this is an empirical determination method. More advanced physics-based methods for determining wake width\cite{13vahidi2022physics,14cheng2018simple,15du2022physical} can also be used in the future.

To maintain the continuity of the Super-Gaussian exponent at the boundary between the near-wake and far-wake regions, we first define $n(x_{th}) = 2$. Furthermore, we define $n(x<x_0) = 2 + A$ within the pressure recovery point $x_0$, where $A = 4$ is the smoothing coefficient of the Super-Gaussian profile. From $x_0$ to $x_{th}$, we can construct a transition function following Vahidi \& Port{\'e}-Agel\cite{13vahidi2022physics}. Finally, we can obtain the piecewise function for the Super-Gaussian exponent $n(x)$:
\begin{equation}\label{eq18}
  n(x)=
\begin{cases}
2+A,x<x_0\\
2+A\text{erfc}\left(2\frac{x-x_0}{x_{th}-x_0}\right),x_0< x\leq x_{th}\\
2, x_{th}<x
\end{cases}
\end{equation}

Furthermore, it is reasonable to assume that the velocity deficit within the expansion region satisfies the one-dimensional momentum theory, and therefore the normalized maximum velocity deficit $C_0$ is $1 - \sqrt{1 - C_T}$. By substituting the Super-Gaussian velocity deficit into the integral equation of the momentum conservation theory
\begin{equation}\label{eq19}
  2\pi\rho\int_0^\infty \Delta U(U_0-\Delta U)rdr=\frac{1}{2}\rho C_T\frac{\pi D^2}{4}U_0^2
\end{equation}

we can have:

\begin{equation}\label{eq20}
  C^2(x)-2^{2/n(x)}C(x)+\frac{n(x)C_T}{16\Gamma(2/n(x))\sigma^{4/n(x)}}=0
\end{equation}

where $\Gamma$ is the Gamma function.Therefore, in the near-wake region $(x < x_{th})$, we have
\begin{equation}\label{eq21}
  \sigma(x)=\left(\frac{n(x)C_T}{2^{2/n(x)}C(x)-C^2(x)}\frac{1}{16\Gamma(2/n(x))}\right)^{(n(x)/4)}
\end{equation}

Finally, we can obtain piecewise functions for the normalized maximum velocity deficit $C(x)$ and the wake width $\sigma(x)$, respectively.
\begin{equation}\label{eq22}
  C(x)=
\begin{cases}
1-\sqrt{1-C_T},x<x_0\\
\frac{x-x_0}{x_{th}-x_0}(C_{th}-C_0)+C_0,x_0< x\leq x_{th}\\
1-\sqrt{1-\frac{C_T}{8(\sigma/D)^2}}, x_{th}<x
\end{cases}
\end{equation}
\begin{equation}\label{eq23}
\sigma(x)/D=
\begin{cases}
\left(\frac{n(x)C_T}{2^{2/n(x)}C(x)-C^2(x)}\frac{1}{16\Gamma(2/n(x))}\right)^{(n(x)/4)},0< x\leq x_{th}\\
k^*\frac{x}{D}+\varepsilon, x_{th}<x
\end{cases}
\end{equation}

To propose the piecewise function for $C(x)$, we assume that the normalized maximum wake velocity deficit in the near-wake region varies linearly from $C_0$ to $C_{th}$. Based on these parameters, we can determine the spatial distribution of the wake velocity deficit, and the corresponding radial wake velocity gradient can be analytically expressed as:

\begin{equation}\label{eq24}
  U_\rho(x,r)=\frac{\partial U}{\partial r}=\frac{\partial U_0\left[1-C(x)\exp\left(-\frac{r^{n(x)}}{2\sigma^2(x)}\right)\right]}{\partial r}=U_0\left[C(x)\exp\left(-\frac{r^{n(x)}}{2\sigma^2(x)}\right)\frac{n(x)r^{n(x)-1}}{2\sigma^2(x)}\right]
\end{equation}

\begin{figure}
\centering
\includegraphics[width=0.8\textwidth]{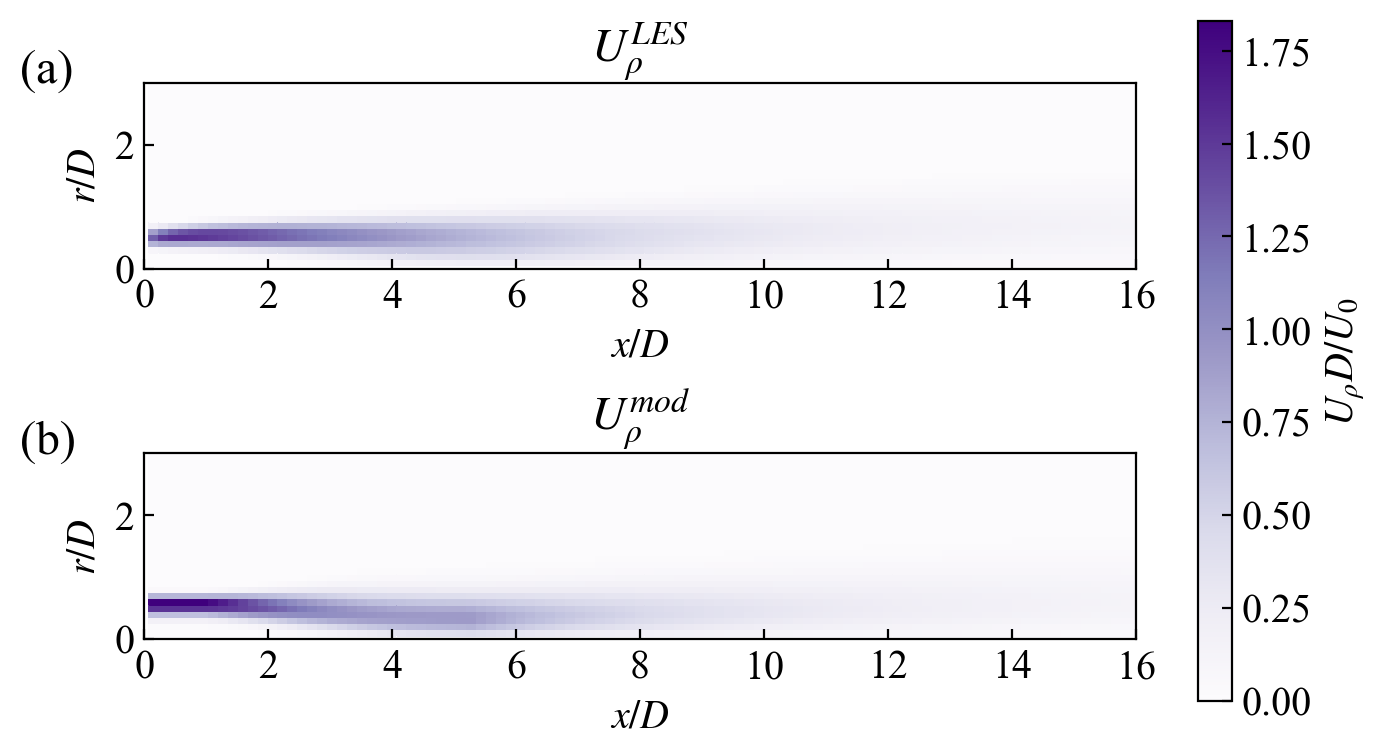}
\caption{Comparison of the normalized radial velocity gradient at hub height obtained from (a) the LES results and (b) the analytical expression for the NBL-3 case.}\label{fig7}
\end{figure}

Figure \ref{fig7} compares the radial velocity gradient obtained from the LES results with the analytical expression for the NBL-3 case. It can be seen that the adopted Super-Gaussian wake model captures the high-velocity shear of the shear layer in the near-wake region very well. The magnitude and spatial distribution of the velocity gradient predicted by the analytical model are in good agreement with the LES results. The results of other cases are similar to this case, so they are not shown here.

\begin{figure}
\centering
\includegraphics[width=0.5\textwidth]{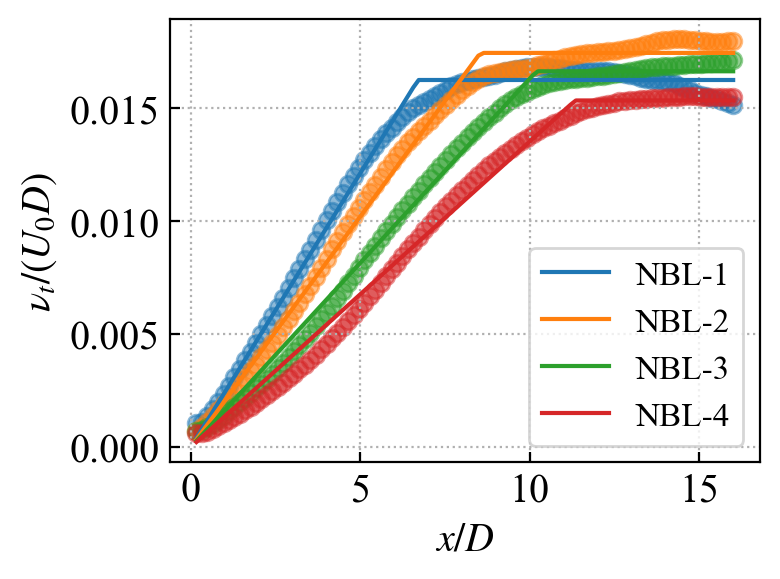}
\caption{Streamwise variation of the turbulent viscosity coefficient $\nu_t(x)$ obtained from curve fitting for all the calibration cases.}\label{fig8}
\end{figure}

\begin{table}
\centering
\renewcommand{\arraystretch}{1.25}
\caption{Overview of the curve fitting parameters for turbulent viscosity coefficient and $\Psi$ for all the calibration cases}\label{tab2}
\begin{tabular*}{0.6\linewidth}{@{}lcccccc@{}}
\toprule
Calibration case & $TI$ & $k_{\nu_t}$ & $x_{\nu_t}$ & $k_{l_m}$
& $k_{C_\varepsilon}$ & $\langle c\rangle_x$ \\
\midrule
NBL-1 & 0.071 & 2.42E-3 & 6.70 & 0.029 & 0.038 & 0.68 \\
NBL-2 & 0.064 & 2.04E-3 & 8.54 & 0.027 & 0.037 & 0.72 \\
NBL-3 & 0.047 & 1.62E-3 & 10.20 & 0.026 & 0.030 & 0.66 \\
NBL-4 & 0.041 & 1.35E-3 & 11.34 & 0.023 & 0.024 & 0.63 \\
\bottomrule
\end{tabular*}
\end{table}
\subsubsection{Turbulent viscosity coefficient $\nu_t$}\label{sec332}

The turbulent viscosity coefficient $\nu_t$ in wind-turbine wakes is a key physical quantity for understanding the turbulent transport process between the wake and ambient flow. It plays an important role in constructing the turbulent transport, dissipation, and shear production terms of the modeled wake-added TKE budget. To obtain $\nu_t$, we fit $-\overline{u^\prime v^\prime}$ at hub height in the main domain to the corresponding radial velocity gradient $\frac{\partial U}{\partial r}$. The fitted values of $\nu_t(x)$ for different calibration cases are shown as points in figure \ref{fig8}. It can be seen that the turbulent viscosity coefficient initially increases approximately linearly and then gradually plateaus. These findings are is in close agreement with those from the wind tunnel experiments \cite{30bastankhah2024modelling,54zong2020point} and numerical simulations \cite{55li2024resolvent,56feng2024resolvent}. Therefore, we can assume that the normalized turbulent viscosity coefficient $\nu_t/(U_0D)$ in wind-turbine wakes takes the form of a piecewise linear function \cite{57klemmer2025wake}:
\begin{equation}\label{eq25}
  \frac{\nu_t(x)}{U_0D}=\begin{cases}
k_{v_t}x/D,\qquad x/D\leq x_{\nu_t}\\
k_{v_t}x_{\nu_t}    ,\qquad  x/D>x_{\nu_t}
\end{cases}
\end{equation}

where $k_{\nu_t}$ is the slope of $\nu_t$ and $x_{\nu_t}$ denotes the plateau point. By fitting the turbulent viscosity coefficient points for different cases using Eq. (\ref{eq25}), the corresponding parameters for the calibration cases can be obtained, as shown in Table \ref{tab2}. We can see that $k_{\nu_t}$ and $x_{\nu_t}$ are closely related to the total turbulence intensity. As the total turbulence intensity increases, $k_{\nu_t}$ becomes larger, and $x_{\nu_t}$ extends further. This variation aligns with the findings of Li et al. \cite{55li2024resolvent}.

\begin{figure}
\centering
\includegraphics[width=0.8\textwidth]{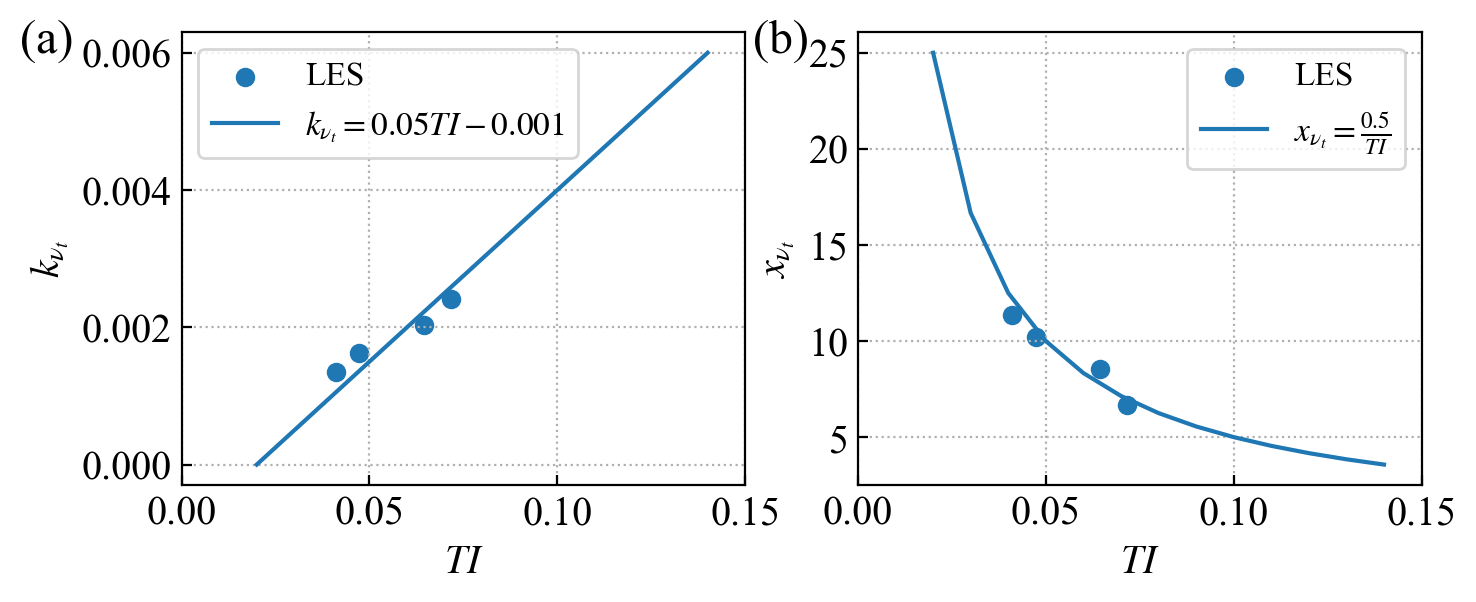}
\caption{(a) Relationship between $k_{\nu_t}$ and total turbulence intensity and (b) relationship between  $x_{\nu_t}$ and total turbulence intensity for all the calibration cases.}\label{fig9}
\end{figure}

Figure \ref{fig9} further illustrates the variation of $k_{\nu_t}$ and $x_{\nu_t}$ with total turbulence intensity $TI$. It can be seen that $k_{\nu_t}$ varies approximately linearly with $TI$, while $x_{\nu_t}$ varies approximately inversely with $TI$. Therefore, the empirical relation for $k_{\nu_t}$ and $x_{\nu_t}$ can be obtained through simple curve fitting, which can then be used to predict the streamwise variation of $\nu_t(x)$ under different inflow conditions.

Finally, we present the empirical expression for determining the turbulent viscosity coefficient in wind-turbine wakes as follows:

\begin{equation}\label{eq26}
  \frac{\nu_t(x)}{U_0D}=\begin{cases}
(0.05TI-0.001)x/D,\qquad x/D\leq 0.5/TI\\
(0.05TI-0.001)\times0.5/TI    ,\qquad  x/D>0.5/TI
\end{cases}
\end{equation}

\subsubsection{Parameter $\Psi$}\label{sec333}

\begin{figure}
\centering
\includegraphics[width=1.0\textwidth]{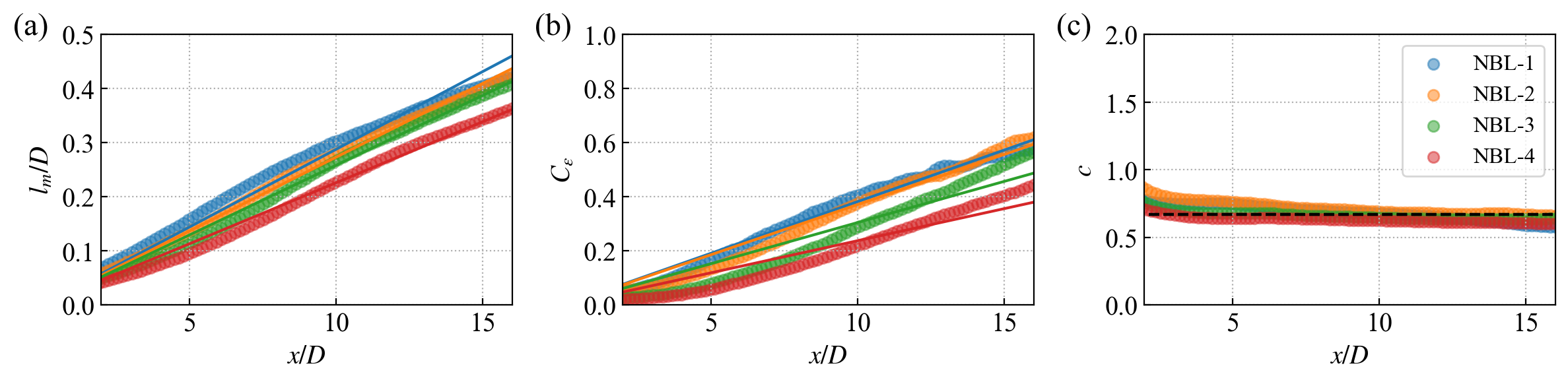}
\caption{Streamwise variation of (a) the normalized mixing length $l_m/D$, (b) the normalized TKE dissipation rate $C_\varepsilon$, and (c) parameter $c$ obtained from curve fitting for all calibration cases.}\label{fig10}
\end{figure}

According to the definition, $\Psi=cl_m^2/C_\varepsilon$. The calculation methods of $l_m$, $C_\varepsilon$ and $c$ are detailed in Sec. \ref{sec32}. By curve fitting, we can obtain $l_m$,$C_\varepsilon$, $c$ for each calibration case, as shown in figure \ref{fig10}. It can be observed that $l_m$ and $C_\varepsilon$ increase quasi-linearly in the streamwise direction, which is consistent with the results obtained by Bastankhah et al. \cite{30bastankhah2024modelling} in wind tunnel experiments. Furthermore, the greater the inflow turbulence, the higher the slope. In contrast, the value of $c$ remains almost constant in the streamwise direction across different cases and can thus can be represented by a constant. Therefore, it is reasonable to assume that the normalized parameters $l_m/D$ and $C_\varepsilon$ take the form of linear functions, that is:

\begin{equation}\label{eq27}
  l_m/D=k_{l_m}x/D\qquad C_\varepsilon=k_{C_\varepsilon}x/D
\end{equation}

where $k_{l_m}$ and $k_{C_\varepsilon}$ represent the slopes of the normalized mixing length and TKE dissipation rate, respectively. By fitting them using Eq. (\ref{eq27}), the corresponding parameters for the calibration cases can be obtained, as shown in Table \ref{tab2}. In addition, table \ref{tab2} also shows the streamwise-averaged parameter $\langle c\rangle_x$. It can be seen that $\langle c\rangle_x$ does not vary significantly across different cases. Therefore, the mean value of 0.67 from these cases is used to model $c$.

\begin{figure}
\centering
\includegraphics[width=0.8\textwidth]{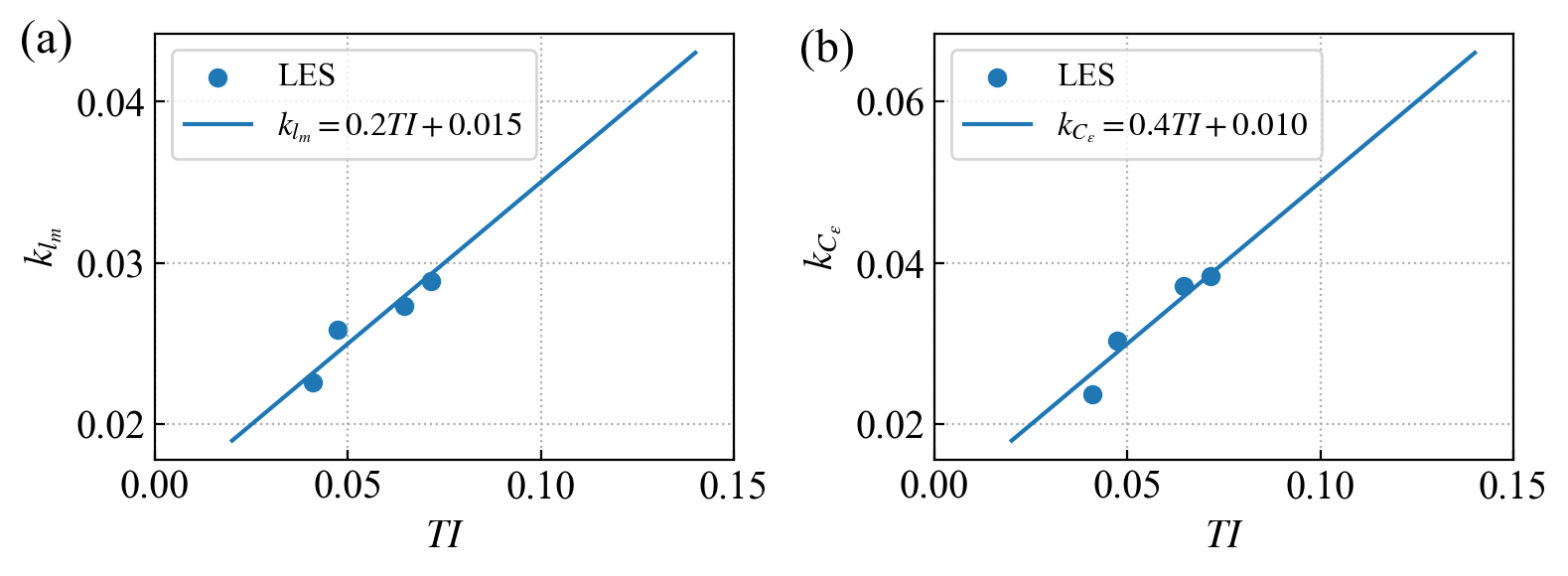}
\caption{(a) Relationship between $k_{l_m}$ and total turbulence intensity and (b) relationship between  $k_{C_\varepsilon}$ and total turbulence intensity for all calibration cases.}\label{fig11}
\end{figure}

Figure \ref{fig11} further shows the variation of  $k_{l_m}$ and $k_{C_\varepsilon}$ with total turbulence intensity $TI$. It can be seen that $k_{l_m}$ and $k_{C_\varepsilon}$ are linear functions of total turbulence intensity. Therefore, the empirical prediction model for $k_{l_m}$ and $k_{C_\varepsilon}$ can be obtained through curve fitting, as shown by the lines in figure \ref{fig11}.

Finally, we obtain the empirical expression for the parameter $\Psi$:
\begin{equation}\label{eq28}
  \frac{\Psi}{D^2}=\frac{cl_m^2}{D^2C_\varepsilon}=0.67\times\frac{(0.2TI+0.015)^2}{0.4TI+0.010}\frac{x}{D}
\end{equation}

The linear variation of the parameter $\Psi$ with downstream distance is consistent with the wind tunnel experimental results of Bastankhah et al. \cite{30bastankhah2024modelling}.

After obtaining the radial wake velocity gradient $U_\rho$, the turbulent viscosity coefficient $\nu_t$, and the key parameter $\Psi$, we can use Eq. (\ref{eq14}) to calculate the azimuthally-averaged wake-added TKE $\langle k_w \rangle_{\theta}(x,r)$, given the inflow wind velocity, turbulence intensity and the thrust coefficient of wind turbine. 

%% file: sections/section_4.tex
\section{Modeling of the ground effect correction term}\label{sec4}

In this section, we propose an analytical expression for the ground effect correction term and provide the determination method for the free parameters based on the calibration cases. Finally, we achieve the accurate modeling of the ground effect correction term based on the inflow ambient information and wind turbine operating conditions.

\subsection{The general function form of the ground effect correction term $\delta_{k_w}(x,r,\theta)$}\label{sec41}

\begin{figure}
\centering
\includegraphics[width=1.0\textwidth]{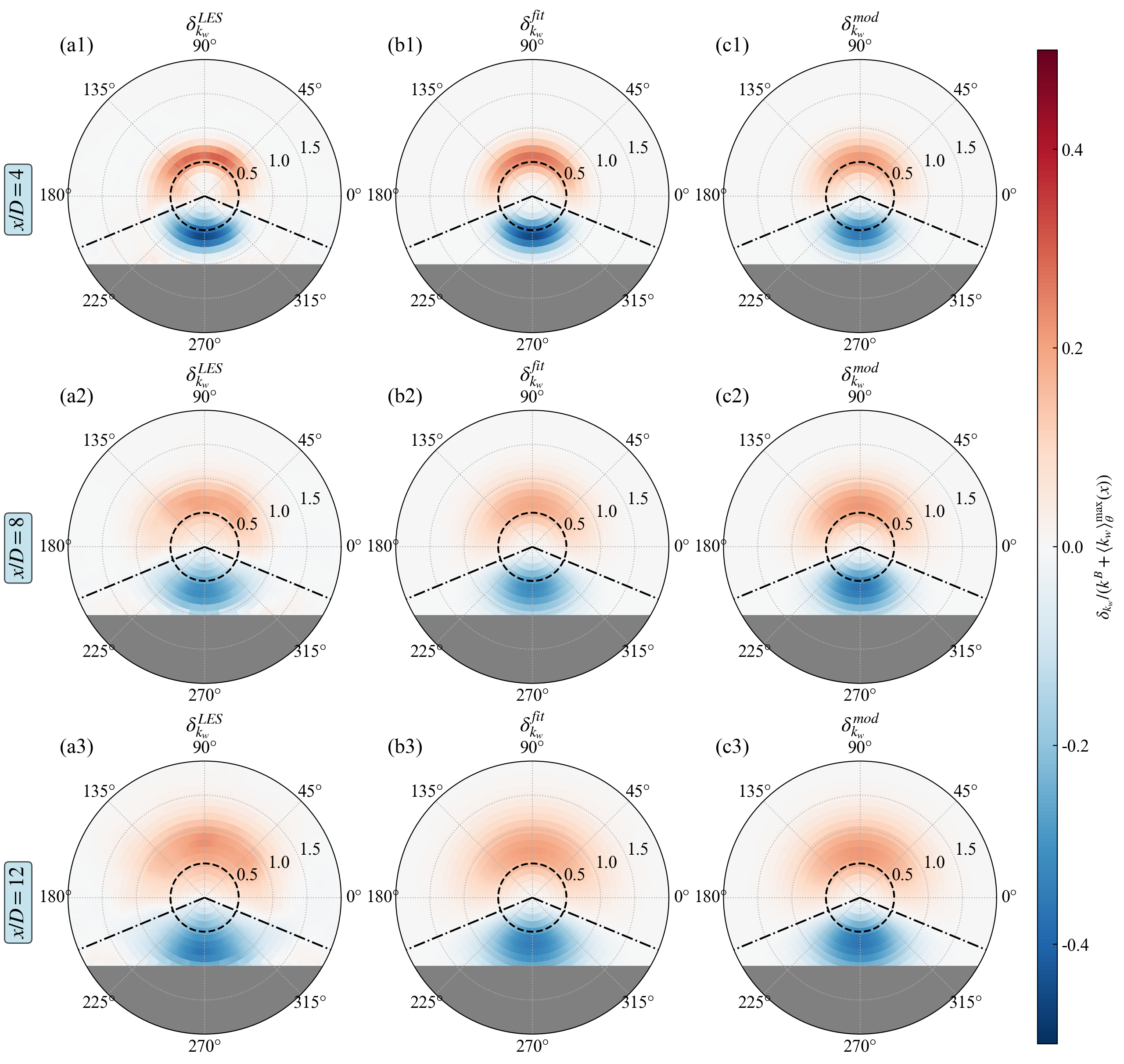}
\caption{Comparison of the ground effect correction function obtained from (a1-a3) LES results, (b1-b3) the corresponding fitted results using Eq. (\ref{eq29}), and (c1-c3) model prediction at different downstream locations of the wind turbine for the NBL-3 case.}\label{fig12}
\end{figure}

As mentioned in Sec. \ref{sec23}, the spatial distribution of wake-added TKE is very similar to that of its dominant term, the wake-added streamwise Reynolds normal stress. Therefore, it is reasonable to employ the functional form of the ground effect correction function for wake-added streamwise turbulence intensity, as defined by Li et al. \cite{18li2022novel}, to the normalized ground effect correction term of wake-added TKE. Unlike Li et al. \cite{18li2022novel}, we redefine the azimuthal range of the ground effect correction function to be greater than 0 and less than 0, so that it more closely matches the realistic spatial distribution characteristics, as shown in figure \ref{fig12} (a1-a3). The ground effect correction model of Li et al. \cite{18li2022novel} assumes that the function is greater than 0 in the region $\theta\in(0,\pi)$, and less than 0 in the region $\theta\in(\pi,2\pi)$. However, the LES results in this study show that the boundary between the region greater than 0 and less than 0 does not appear at hub height, but rather at approximately $\theta=-\pi/8$ and $\theta=9\pi/8$, as shown by the dot-dashed line in figure \ref{fig12}. Therefore, the proposed ground effect correction function has the following expression:
\begin{equation}\label{eq29}
  \frac{\delta_{kw}(x,r,\theta)}{k^B+\langle k_w\rangle_\theta^{\max}(x)}=
\begin{cases}
B\sin\left((\theta+\frac{\pi}{8})\frac{4}{5}\right)\left[k_1(x,r)\exp\left(-\frac{(r-r_{\delta}(x))^2}{2\sigma_\delta(x)^2}\right)\right],(-\frac{\pi}{8}\le\theta\le \frac{9\pi}{8})\\
C\sin\left((\theta-\frac{9\pi}{8})\frac{4}{3}+\pi\right)\left[k_1(x,r)\exp\left(-\frac{(r-r_{\delta}(x))^2}{2\sigma_\delta(x)^2}\right)\right],(\frac{9\pi}{8}<\theta< \frac{15\pi}{8})
\end{cases}
\end{equation}

where $\langle k_w\rangle_\theta^{\max}(x)$ is the maximum azimuthally-averaged wake-added TKE at different streamwise locations, $r_{\delta}(x)$ is the radial position of the maximum of $\delta_{k_w}$, $\sigma_\delta(x)$ is the characteristic width of $\delta_{k_w}$, and $B$ and $C$ are the empirical coefficients corresponding to the upper and lower half-sectors of $\delta_{k_w}$, respectively. $k_1(x,r)$ is the shape factor, and it can be calculated using the following expression:

\begin{equation}\label{eq30}
  k_1(x,r)=
\begin{cases}
\sin\left(\frac{\pi}{2}\frac{r}{r_{\delta}(x)}\right), (r<r_{\delta}(x))\\
1, (r\ge r_{\delta}(x))
\end{cases}
\end{equation}

Figure \ref{fig12} shows the comparison between the ground effect correction function obtained from LES results and the corresponding fitted results using Eq. (\ref{eq29}) at different locations downstream of the wind turbine for the NBL-3 case. It can be seen that, with the fitted parameters, the proposed functional form can accurately predict the spatial distribution of the ground effect correction function. The results of other cases are similar to this case, so they are not shown here.

\subsection{Determination of the key parameters}\label{sec42}

To calculate $\delta_{k_w}(x,r,\theta)$, which is a component of the wake-added TKE, it is necessary to determine its maximum radial position $r_\delta(x)$, its characteristic width $\sigma_\delta(x)$, and its empirical coefficients $B$ and $C$. Among them, $r_\delta(x)$ and $\sigma_\delta(x)$ mainly affect the influence extent of $\delta_{k_w}(x,r,\theta)$, while the empirical correction coefficients $B$ and $C$ mainly influence its magnitude. According to the definition of $\delta_{k_w}(x,r,\theta)$, we have $\frac{1}{2\pi}\int_0^{2\pi}\delta_{k_w}(x,r,\theta)d\theta=0$, and therefore $C=\frac{5}{3}B$. $r_\delta(x)$, $\sigma_\delta(x)$, and $B$ can be determined by fitting the LES results from different calibration cases.
\begin{figure}
\centering
\includegraphics[width=1.0\textwidth]{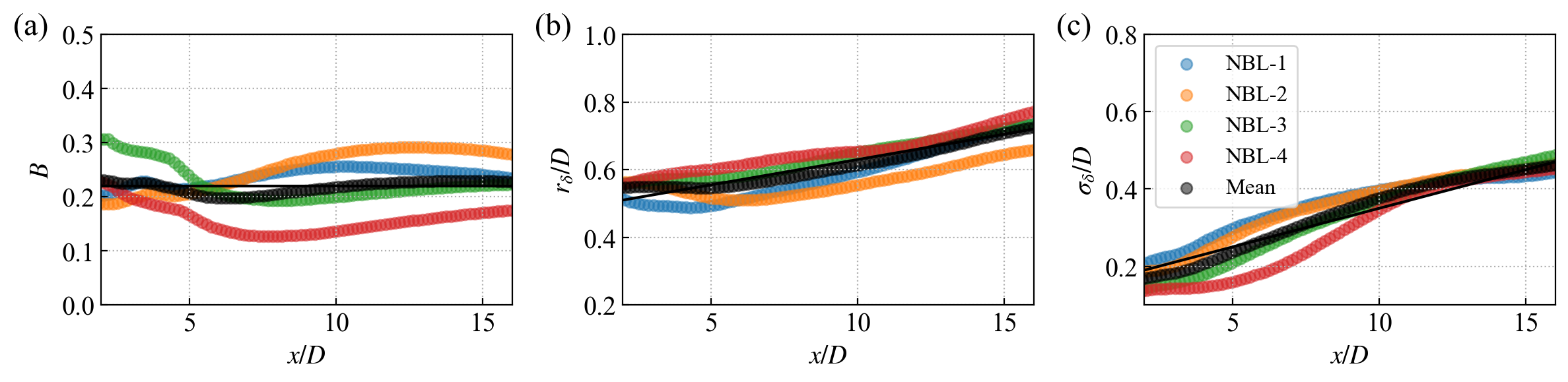}
\caption{Streamwise variation of (a) the empirical correction coefficients $B$, (b) the maximum radial position $r_\delta(x)$, and (c) the characteristic width $\sigma_\delta(x)$ of the ground effect correction function obtained from fitting using Eq. (\ref{eq29}) for all calibration cases.}\label{fig13}
\end{figure}

As can be seen, $r_\delta(x)$ and $\sigma_\delta(x)$ in different cases exhibit quasi-linear growth in different cases, while the streamwise variation of the empirical correction coefficient $B$ remains relatively steady. The small differences in their magnitudes across different cases indicate that the ground effect correction function exhibits similarity. Therefore, the averaged value across all calibration cases can be calculated, as shown by the black points in figure \ref{fig13}. $r_\delta(x)$ and $\sigma_\delta(x)$ can be approximated by linear functions, while $B$ can be approximated by a constant function. Although the corresponding parameters slightly vary across different cases, the model validation results in Sec. \ref{sec5} show that using the mean value does not lead to significant errors. In summary, the free parameters in the ground effect correction function can be determined using the following expressions, as shown by the black solid lines in figure \ref{fig13}.

\begin{equation}\label{eq31}
  B=0.22\qquad\frac{r_\delta}{D}= 0.015\frac{x}{D}+0.48 \qquad\frac{\sigma_\delta}{D}= 0.02\frac{x}{D}+0.15
\end{equation}

After obtaining the parameters above, the ground effect correction function $\delta_{k_w}(x,r,\theta)$ can be calculated using Eq. (\ref{eq29}). By substituting the above empirical parameters into Eq. (\ref{eq29}), the predicted ground effect correction function for the NBL-3 case is shown in figure \ref{fig12} (c1-c3). It can be seen that the proposed model parameters effectively predict the spatial distribution of the ground effect correction function, providing a solid foundation for accurately predicting the wake-added TKE. 

%% file: sections/section_5.tex
\section{Model validation}\label{sec5}

Upon obtaining the two submodels for the azimuthally-averaged wake-added TKE $\langle k_w\rangle_\theta(x,r)$ and the ground effect correction function $\delta_{k_w}(x,r,\theta)$, the three-dimensional wake-added TKE $k_w$ in wind-turbine wakes can be predicted as follows:
\begin{equation}\label{eq32}
  k_w(x,r,\theta)=\langle k_w\rangle_\theta(x,r)+\delta_{k_w}(x,r,\theta)
\end{equation}

In this section, we evaluate the performance of the proposed model by comparing it with the LES calibration cases and other publicly available wake-added TKE datasets.

\subsection{Comparison with the LES calibration cases}\label{sec51}

\begin{figure}
\centering
\includegraphics[width=1.0\textwidth]{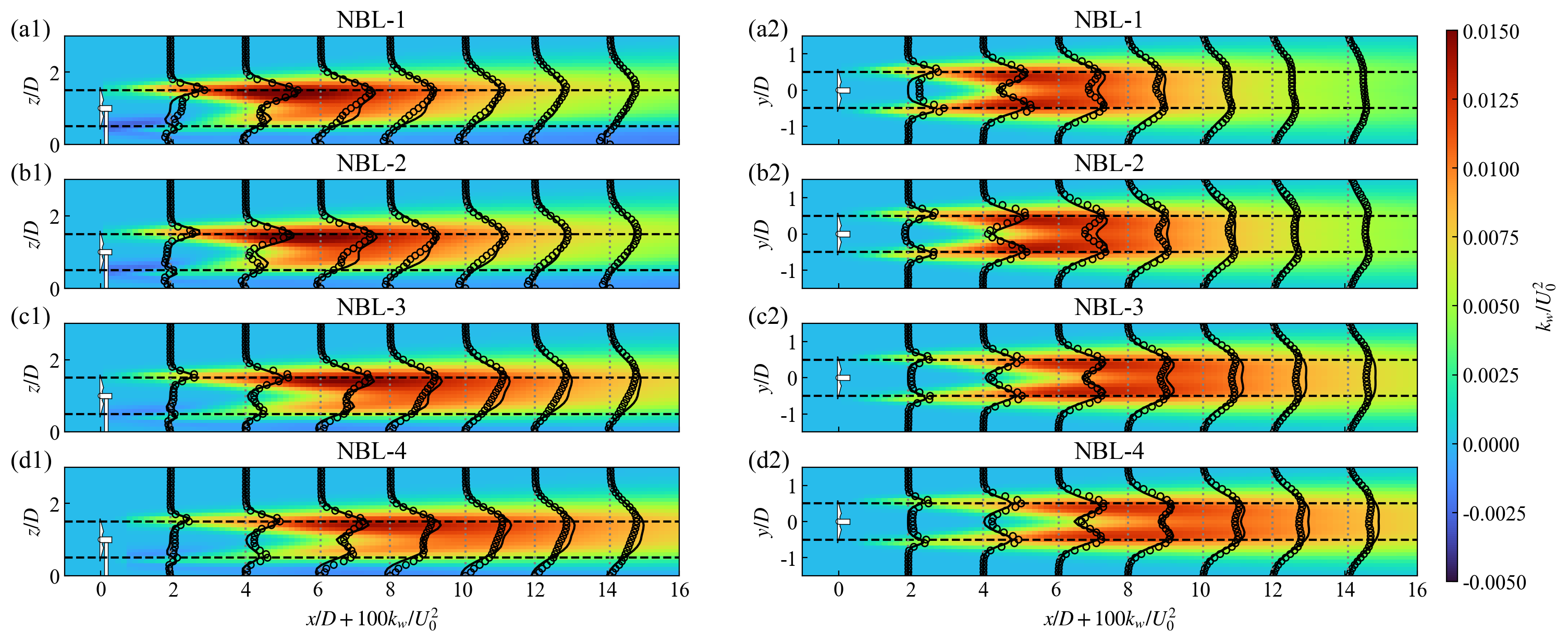}
\caption{Contourfs of the wake-added TKE at hub height and the vertical plane across the rotor center predicted by the proposed model for all calibration cases. The gray dotted lines in each subplot represent the sampling lines of wake-added TKE downstream of the wind turbine. The black solid line and black circle represent the model prediction results and LES results at the sampling lines, respectively. The black dashed lines are the boundary of the rotor swept region.}\label{fig14}
\end{figure}

Figure \ref{fig14} shows the contourfs of the wake-added TKE at hub height and the vertical plane across the rotor center, as predicted by the proposed model for all calibration cases. As expected, the proposed model accurately predicts the three-dimensional spatial distribution of wake-added TKE downstream of the wind turbine, based on inflow information and turbine operation conditions. Interestingly, the model effectively captures the phenomenon that as ambient turbulence increases from the NBL-4 case to the NBL-1 case, the streamwise location of the maximum wake-added TKE and the merge of the dual-peak profile advance, while the decay of wake-added TKE becomes more rapid. This comparison preliminarily demonstrates the validity of the physics-based azimuthally-averaged wake-added TKE prediction model and the proposed parameter determination method. Furthermore, the proposed model accurately captures the influence of inflow velocity shear on the asymmetry of wake-added TKE vertical profiles, where the wake-added TKE at the top-tip height is significantly higher than that at the lower-tip height, thus preliminarily validating the proposed ground effect correction function and its parameter determination method.

\subsection{Comparison with the public datasets from the literature}\label{sec52}

As mentioned in the introduction, wind energy researchers primarily focused on the wake-added streamwise turbulence intensity in the early days \cite{16crespo1996turbulence,17ishihara2018new,18li2022novel,19tian2022new}. Since it is relatively easy to measure, there are numerous field observations, wind tunnel experiments, and numerical simulation datasets\cite{4porte2020wind}. In contrast, high-quality datasets of wake-added TKE remain relatively scarce. Nevertheless, we collected as much wake-added TKE data as possible from the literature to validate the prediction accuracy of the proposed model, as shown in table \ref{tab3}. The validation cases include LES and wind tunnel experiments, covering a range of inflow wind velocities, turbulence intensities, turbine geometries, and thrust coefficients. To quantify the prediction error of the proposed model, we defined the normalized mean absolute error ($NMAE$) for each case as follows:
\begin{equation}\label{eq33}
  NMAE=\frac{\frac{1}{N}\sum_i^N|k_w^{obs,i}-k_w^{mod,i}|}{k^B+\max(k_w^{obs})}\times 100\%
\end{equation}

where $N$ is the total number of observation points in each case, $|\cdot|$ represents the absolute value, the superscripts $^{obs}$ and $^{mod}$ represent the observed and predicted values, respectively, $i$ is the index of the observation point, and $\max(\cdot)$ denotes the maximum value.

\begin{table}
\renewcommand{\arraystretch}{1.25}
\caption{Overview of the validation datasets}\label{tab3}
\begin{tabular*}{1.0\linewidth}{@{}lcccccccc@{}}
\toprule
Case & Type & \makecell[c]{Rotor \\diameter \\ $D$(m)} & \makecell[c]{Hub \\height \\ $H$(m)} & \makecell[c]{Mean wind\\ velocity at\\ hub height\\ $U_0$(m/s)} & \makecell[c]{Thrust\\ coefficient\\ $C_T$} &\makecell[c]{Total\\ turbulence\\intensity at\\ hub height\\ $TI$} & \makecell[c]{Streamwise\\ turbulence\\intensity at\\ hub height\\ $I_u$} & $NMAE$\\
\midrule
BLO2025-SBL & LES & 0.15 & 0.125 & 2.5 & 0.8 & 0.04 & 0.05 & 5.5\% \\
BLO2025-RBL & LES & 0.15 & 0.125 & 2.5 & 0.8 & 0.078 & 0.10 & 3.1\% \\
WU2023 & LES & 77 & 80 & 9.16 & 0.68 & 0.0625 & 0.08 & 9.6\% \\
ARC2020 & LES & 126 & 90 & 9.00 & 0.83 & 0.080 & 0.102 & 8.7\% \\
AJU2020 & Wind tunnel & 0.2 & 0.2 & 6.43 & 0.585 & 0.094 & 0.120 &
9.9\% \\
BAS2024 & Wind tunnel & 0.416 & 0.30 & 1.42 & 0.48 & 0.048 & 0.061 &
12.0\% \\
\bottomrule
\end{tabular*}
\end{table}

\begin{figure}
\centering
\includegraphics[width=1.0\textwidth]{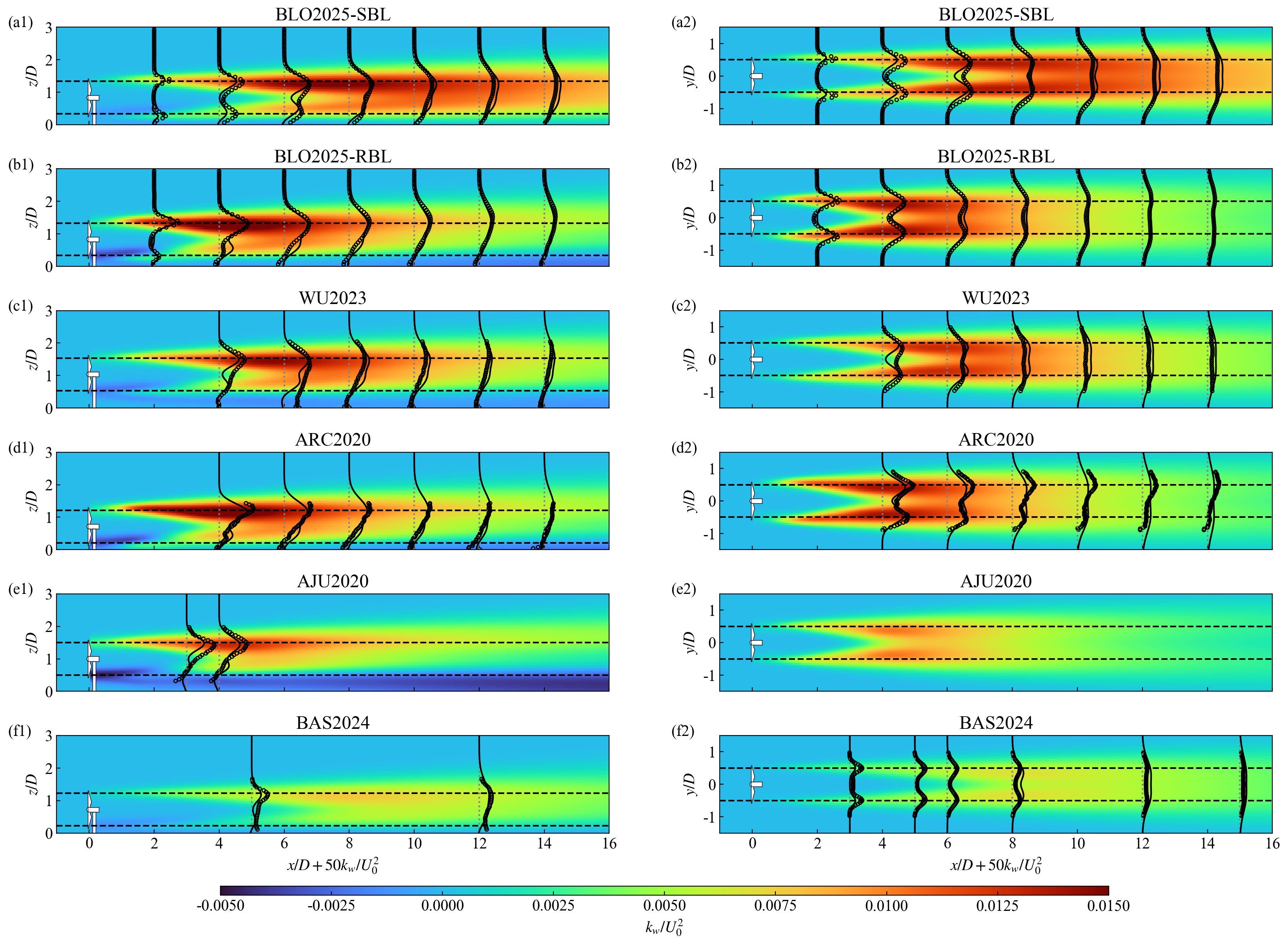}
\caption{Contourfs of the wake-added TKE at hub height and the vertical plane across the rotor center predicted by the proposed model for all validation cases. The gray dotted lines in each subplot represent the sampling lines of wake-added TKE downstream of the wind turbine. The black solid line and black circle represent the model prediction results and LES results at the sampling lines, respectively. The black dashed lines are the boundary of the rotor swept region.}\label{fig15}
\end{figure}
The BLO2025-SBL and BLO2025-RBL cases are two stand-alone wind turbine LES cases under the neutral boundary layer, simulated using the waLBerla-wind LBM-LES solver by Blondel et al. \cite{31blondel2025physics}. The simulated wind turbine has a rotor diameter of 0.15m, a hub height of about 0.125m, and a thrust coefficient of 0.8. The simulated mean wind speed at hub height is 2.5m/s. The streamwise turbulence intensity at hub height is 0.05 and 0.10, with corresponding total turbulence intensity of 0.04 and 0.78, respectively. The vertical wake-added TKE profiles across the rotor center and the spanwise hub-height wake-added TKE profiles at distances of $2D$, $4D$, $6D$, $8D$, $10D$, $12D$, and $14D$ downstream of the wind turbine are used to validate the performance of the proposed model for these two cases, as shown by the points in figure \ref{fig15} (a1-b2).

Similarly, the WU2023 and ARC2020 cases are two utility-scale wind turbine LES cases under the neutral boundary layer, performed using the WRF-LES model and the open-source SOWFA tool by Wu et al. \cite{58wu2024new} and Archer et al. \cite{59archer2020two}, respectively. In the WU2023 case, the simulated wind turbine is the PSU1.5MW wind turbine, with a rotor diameter of 80m, a hub height of 77m, and a thrust coefficient of 0.68. The surface roughness is 0.01m, and the corresponding mean wind speed at hub height is 9.16m/s. The streamwise turbulence intensity at hub height is 0.08, with the corresponding total turbulence intensity approximately 0.0625, as determined by the empirical expression $I_u=1.28TI$. In the ARC2020 case, the simulated wind turbine is the NREL-5MW wind turbine, with a rotor diameter of 126m and a hub height of 90m. The simulated mean wind speed at hub height is 9.00m/s, and the streamwise turbulence intensity at hub height is 0.102. According to the empirical expression, the total turbulence intensity is approximately 0.08. Khanjari et al. \cite{29khanjari2025analytical} provide the vertical wake-added TKE profiles across the rotor center and the spanwise hub-height wake-added TKE profiles at distances of $4D$, $6D$, $8D$, $10D$, $12D$, and $14D$ downstream of the wind turbine for these two cases, as shown by the points in figure \ref{fig15} (c1-d2).

The AJU2020 case refers to the wind tunnel experiments performed by Aju et al. \cite{60aju2020influence}. The rotor diameter and hub height of the wind turbine used in this experiment are both 0.2m, and the thrust coefficient is 0.585. The corresponding mean wind speed at hub height is 6.43m/s, and the streamwise turbulence intensity at hub height is 0.120. According to the empirical expression, the total turbulence intensity is estimated to be 0.094. Khanjari et al. \cite{29khanjari2025analytical} provides the vertical wake-added TKE profiles across the rotor center at $3D$ and $4D$ downstream of the wind turbine for the AJU2020 case, as shown by the points in figure \ref{fig15} (e1).

The BAS2024 case refers to the wind tunnel experiments conducted by Bastankhah et al.\cite{30bastankhah2024modelling}. The small wind turbine used in this experiment has a rotor diameter of 0.416m, a hub height of 0.30m, and a thrust coefficient of 0.48. The corresponding mean wind speed at hub height is 1.42m/s, and the total turbulence intensity at hub height is 0.048. According to the empirical relation, the streamwise turbulence intensity at hub height is approximately 0.061. Bastankhah et al. \cite{30bastankhah2024modelling} provided the vertical TKE profiles across the rotor center upstream, at $5D$ and $12D$ downstream of the wind turbine in this case. These data can be used to calculate the wake-added TKE vertical profiles at $5D$ and $12D$ downstream of the wind turbine, as shown by the points in figure \ref{fig15}(f1). In addition, they presented the radial profiles of the hub-height wake-added TKE at distances of $3D$, $5D$, $6D$, $8D$, $12D$, and $15D$ downstream of the wind turbine, as shown by the points in figure \ref{fig15}(f2).

Figure \ref{fig15} shows the contourfs of the wake-added TKE at hub height and the vertical plane across the rotor center, as predicted by the proposed model for all validation cases. In addition, the prediction results of the proposed model at the observation points are shown by the black solid line in figure \ref{fig15}. Overall, the proposed model accurately predicts the spatial distribution characteristics of wake-added TKE in both the vertical and spanwise directions for all validation cases well. Specifically, the spanwise wake-added TKE profile at hub height transitions from a dual-peak distribution in the near-wake region to an approximately top-hat distribution in the far-wake region. The peak value of the wake-added TKE at the top-tip height is significantly higher than that at the lower-tip height. Furthermore, the proposed model satisfactorily captures the influence of the thrust coefficient and inflow turbulence on the streamwise evolution of wake-added TKE. In general, the magnitude of wake-added TKE is most sensitive to the thrust coefficient, while the streamwise location of the maximum wake-added TKE and the merge of the dual-peak profile are most sensitive to the inflow turbulence. Quantitatively, the proposed model exhibits some errors in predicting the magnitude of wake-added TKE, which may be attributed to the neglect of the rotor rotational effect, inhomogeneous background flow, and the high velocity shear region caused by the nacelle in the model development. Specifically, the proposed model slightly underestimates wake-added TKE in the wake region where $x \le 6D$, and slightly overestimates it in the further downstream region for the BLO2025-SBL case, the WU2023 case, and the BAS2024 case. Moreover, this overestimation decreases monotonically with increasing streamwise distance.

In summary, the blind test results from the six validation cases demonstrate that the proposed model can effectively capture the spatial distribution of wake-added TKE under varying inflow and wind turbine operating conditions. Specifically, it satisfactorily predicts the streamwise evolution from the dual-peak distribution to the nearly top-hat distribution of wake-added TKE at hub height, as well as the vertical asymmetry of the wake-added TKE caused by the ground. Quantitatively, the $NAMEs$ in the six validation cases are all less than 12.0\%, with an average value of about 8.13\%, further confirming the robustness and broad applicability of the proposed model and parameter determination method. In the future, more high-quality wake-added TKE datasets are required to further validate the prediction accuracy of the proposed model.

%% file: sections/section_6.tex
\section{Conclusion}\label{sec6}
Accurate prediction of TKE in wind-turbine wakes is of significant scientific value for understanding the wake recovery mechanisms. Moreover, this physical quantity is a critical input for engineering applications, such as wake velocity deficit evaluation and fatigue damage assessment of downstream wind turbines. However, the existing empirical model \cite{29khanjari2025analytical}, which assumes a specific profile form, lacks a solid theoretical foundation and exhibits poor physical interpretability. On the other hand, physics-based models face issues such as low prediction accuracy in the near-wake region \cite{31blondel2025physics}, undermined model parameters \cite{30bastankhah2024modelling,31blondel2025physics}, and applicability limited to hub height \cite{30bastankhah2024modelling}, making it impossible to achieve a closed prediction of the three-dimensional spatial distribution of wake-added TKE, given the basic inflow and wind turbine operating conditions. To address these issues, we propose a novel wake-added TKE prediction model capable of accurately predicting the streamwise evolution and three-dimensional spatial distribution of wake-added TKE using the basic inflow (i.e., turbulence intensity and wind speed at hub height) and wind turbine operating conditions (i.e., rotor diameter, hub height, and thrust coefficient). The proposed prediction model consists of two sub-modules: the azimuthally-averaged wake-added TKE calculation module and the ground effect correction function calculation module.

The azimuthally-averaged wake-added TKE $\langle k_w\rangle_{\theta}$ calculation module is based on a rigorous derivation of the TKE transport equation. Specifically, we investigated the wake-added TKE budget in cylindrical coordinates based on LES results, derived a simplified azimuthally-averaged wake-added TKE budget, and modeled the dominant terms within it. Subsequently, we developed the analytical solution for the azimuthally-averaged wake-added TKE, following the approach of Bastankhah et al. \cite{30bastankhah2024modelling}. Compared to Bastankhah et al. \cite{30bastankhah2024modelling}, we proposed methods for determining all the unknown parameters, such as the radial velocity gradient, turbulent viscosity coefficient, mixing length, and normalized TKE dissipation rate, using basic inflow and wind turbine operating conditions. This resulted in a fully-closed calculation method, and achieved end-to-end prediction of the azimuthally-averaged wake-added TKE.

The determination of the ground effect correction function $\delta_{k_w}$ is based on its self-similarity. LES results demonstrate that the ground effect correction function at various downstream locations satisfies self-similarity. Specifically, the correction function follows the Gaussian-like distribution in the radial direction and approximately satisfies the sinusoidal-like distribution in the azimuthal direction. Consequently, it can be represented by a unified functional form. Furthermore, the ground effect correction function also requires the determination of several parameters, including the radial position of the maximum of $\delta_{k_w}$, the characteristic width of $\delta_{k_w}$, and the empirical coefficients for the upper and lower half sectors of $\delta_{k_w}$. These parameters can be obtained through parameter fitting using calibration cases. Finally, we developed empirical expressions for these parameters based on basic inflow and wind turbine operating conditions, enabling end-to-end prediction of the ground effect correction function.

By combining the above two sub-modules, the proposed model enables the prediction of the three-dimensional spatial distribution of wake-added TKE. We compared the proposed model with LES calibration cases and publicly available validation datasets from the literature. The validation datasets consist of LES and wind tunnel experiments, including various inflow and wind turbine operating conditions. The results indicate that the proposed model can satisfactorily predict the spatial distribution of wake-added TKE, particularly in capturing the vertical asymmetry of wake-added TKE and the streamwise evolution of the hub-height wake-added TKE profile from the dual-peak distribution to the approximate top-hat distribution. Quantitatively, the normalized mean absolute error of the proposed model on the validation dataset is less than 12\%, with an average value of about 8.13\%, most of which originated from the neglected effects of rotor rotation, background inhomogeneity, and the nacelle. These results demonstrate the robustness and broad applicability of the proposed model and its corresponding parameter determination methods.

Although progress have made in the modeling of wake-added TKE, further research is needed in several areas. For instance, the proposed model has only been validated under neutral atmospheric conditions, and future work should could focus on validating its applicability under different atmospheric stability conditions. Additionally, in the development of the wake-added TKE model, the influence of wind veer induced by the Coriolis force was neglected, and this effect on wake-added TKE is worthy of further investigation. 